
%
%

\input phyzzx

\def\bt#1{\hbox{$\lower5pt\hbox{$#1$}\atop\sim$}}
\def\sqr#1#2{{\vcenter{\hrule height.#2pt
        \hbox{\vrule width.#2pt height #1pt \kern#1pt
        \vrule width.#2pt}
        \hrule height.#2pt}}}

\def\ri{\rightarrow}
\def\ca{{^{\rm CA}}}
\def\aca{A^{^{\rm CA}}}
\def\pc{{\rm pc}}
\def\pv{{\rm pv}}
\def\H{{\cal H}}
\def\up{\uparrow}
\def\dw{\downarrow}
\def\halfm{{1\over 2}^{_-}}
\def\halfp{{1\over 2}^{_+}}
\def\half{{1\over 2}}
\def\thalfp{{3\over 2}^{_+}}
\def\lamc{{\Lambda_c^{^+}}}
\def\xic{{\Xi_c^0}}
\def\xip{{\Xi_c^{^{+A}}}}
\def\xin{{\Xi_c^{^{0A}}}}
\def\xins{{\Xi_c^{0S}}}
\def\sigz{\Sigma^{^0}}
\def\tx{\tilde{X}}

\def\dup{d^{\uparrow}}
\def\ddw{d^{\downarrow}}
\def\sup{s^{\uparrow}}
\def\sdw{s^{\downarrow}}
\def\cup{c^\uparrow}
\def\cdw{c^\downarrow}

\def\tdup{\tilde{d}^{\uparrow}}
\def\tddw{\tilde{d}^{\downarrow}}
\def\tsup{\tilde{s}^{\uparrow}}

\def\bdup{\bt{d}^\up}
\def\bddw{\bt{d}^\dw}

\def\bdupl{\bt{d}^{^\big{\uparrow}}}

\def\pole{ {^{\rm pole}}}
\def\nf{ {{\rm n.f.}}}
\def\fac{ {{\rm fac}}}
\def\bse{{\bf 70}}
\def\bsi{{\bf 6}}
\def\be{{\bf 8}}
\def\bten{{\bf 10}}
\def\siga{\Sigma^{^+}(\be,P_{1/2})_a}
\def\sigc{\Sigma^{^+}(\be,P_{3/2})}

\def\pr{{\sl Phys. Rev.}~}
\def\prl{{\sl Phys. Rev. Lett.}~}
\def\pl{{\sl Phys. Lett.}~}
\def\np{{\sl Nucl. Phys.}~}
\def\zp{{\sl Z. Phys.}~}

\font\el=cmbx10 scaled \magstep2
{\obeylines
\hfill IP-ASTP-10-93
\hfill ITP-SB-93-20
\hfill April, 1993}

\vskip 0.5 cm

\centerline {{\el Cabibbo-allowed nonleptonic weak decays of charmed baryons}}

\bigskip

\centerline{\bf Hai-Yang Cheng$^{a,b,*}$ and B. Tseng$^{a}$}

\medskip
\centerline{$^a$ Institute of Physics, Academia Sinica}
\centerline{Taipei, Taiwan 11529, Republic of China}

\medskip
\centerline{$^b$ Institute for Theoretical Physics, State University of
New York}
\centerline{Stony Brook, New York 11794, USA}

\medskip
\bigskip

\centerline{\bf Abstract}

  Cabibbo-allowed nonleptonic weak decays of charmed baryons $\lamc,~
\xin,~\xip$ and $\Omega_c^{^0}$ into an octet baryon and a pseudoscalar
meson are analyzed. The nonfactorizable contributions are evaluated under pole
approximation, and it turns out that the $s$-wave amplitudes are dominated by
the low-lying $\halfm$ resonances, while $p$-wave ones governed by the
ground-state $\halfp$ poles.
The MIT bag model is employed to calculate the coupling constants, form
factors and baryon
matrix elements. Our conclusions are: (i) $s$ waves are no longer dominated
by commutator terms; the current-algebra method is certainly not applicable to
parity-violating amplitudes, (ii) nonfactorizable $W$-exchange effects are
generally important; they can be comparable to and somtimes even dominate over
factorizable contributions, depending on the decay modes under consideration,
(iii) large-$N_c$ approximation for factorizable amplitudes also works in
the heavy baryon sector and it accounts for the color nonsuppression of
$\lamc\ri p\bar{K}^0$ relative to $\lamc\ri\Lambda\pi^+$,
(iv) a measurement of the decay rate and the sign of the $\alpha$ asymmetry
parameter of certain proposed decay modes will help discern various models;
especially the sign of $\alpha$ in $\lamc\ri\Sigma\pi$ decays can be used
to unambiguously differentiate recent theoretical schemes from current
algebra, and (v) $p$ waves are the dominant contributions to the decays
$\lamc\ri\Xi^0 K^+$ and $\xin\ri\Sigma^+ K^-$, but they are subject to
a large cancellation; this renders present theoretical predictions on these
two channels unreliable.
\vskip 0.5cm
\noindent $^*$E-mail: phcheng@twnas886.bitnet, cheng@max.physics.sunysb.edu
\endpage

\chapter{Introduction}
   With more and more data of charmed baryon decays becoming available at
ARGUS, CLEO, CERN and Fermilab, it reaches the point that a systematical and
serious theoretical study of the underlying mechanism for nonleptonic
decays of charmed baryons is called for [1]. The experimental progress
in this area is best summarized in the recent concluding remark by
Butler [2] that ``Our knowledge of the charmed baryons has taken another
leap forward. This is a field whose time has finally arrived.'' Indeed, in the
past few years, new and high-statistics measurements of the nonleptonic $\lamc$
decays have been carried out, and new decay modes of $\xin$ and
$\Omega_c$ also have been seen recently.

   Theoretically, all nonleptonic weak decays of mesons and baryons can be
classified in terms of the following quark diagrams [3]:
\foot{The $W$-annihilation diagram is absent in the baryon decay. The $W$-loop
diagram does not contribute to the Cabibbo allowed weak decays of hadrons.}
the external $W$-emission diagram, the internal $W$-emission diagram, the
$W$-exchange diagram, the $W$-annihilation diagram, and the $W$-loop diagram.
The external and internal $W$-emission diagrams are usually referred to as
factorizable contributions.
\foot{In general, there are two distinct internal $W$-emission diagrams and
three $W$-exchange diagrams for the nonleptonic baryon decay [4]. However,
only the internal $W$-emission diagram with the meson formed along the parent
quark which decays weakly is factorizable. At the hadron level, the
factorizable internal $W$-emission graph corresponds to a meson-pole
contribution.}
It is known for meson nonleptonic decays that the factorizable contribution
dominates over the nonfactorizable ones such as $W$-exchange and
$W$-annihilation. For baryon decays, {\it a priori} the nonfactorizable
contribution can be as important as the factorizable one since $W$-exchange,
contrary to the meson case, is no longer subject to helicity and color
suppression.

   How do we handle the $W$-exchange contribution in the baryon decay? In
principle the $W$-exchange amplitude can be expressed as a sum of all possible
intermediate hadronic states. In practice, one assumes pole approximation,
namely that only one-particle intermediate states are kept; that is, the
$W$-exchange contribution is assumed to be approximately saturated by pole
intermediate states. Among all possible pole contributions, including
resonances and continnum states, one usually concentrates on the most important
poles such as
the low-lying $J^P=\halfp,~\halfm$ states. In general, nonfactorizable $s$- and
$p$-wave amplitudes are dominated by $\halfm$ low-lying baryon resonances and
$\halfp$ ground-state baryon poles, respectively. Evidently, the estimate
of the $s$-wave terms is a difficult and nontrivial task since it involves
weak baryon matrix elements and strong coupling constants of $\halfm$
baryon states, which we know very little. Nevertheless, there is one
exceptional case: For hyperon nonleptonic decays, the evalution of $s$ waves
is no more difficult than the $p$-wave amplitudes.
This comes from the fact that the
emitted pion in this case is soft. As a result, the parity-violating pole
amplitude of the hyperon decay is reduced, in the soft pion limit, to the
familiar equal-time commutator terms. The magic feature with this current
algebra approach is that the $s$-wave amplitude can now be manipulated without
appealing to any information of the cumbersome $\halfm$ poles.

   Traditionally, the two-body nonleptonic weak decays of charmed baryons is
studied by utilizing the same technique of current algebra as in the case of
hyperon decays [5-13]. However, the use of the soft-meson theorem makes sense
only if the emitted meson is of the pseudoscalar type and its momentum is
soft enough. Obviously, the pseudoscalar-meson final state in charmed bayon
decay is far from being ``soft''. Therefore, it is not appropriate to
make the soft meson limit. Moreover, since the charmed bayon is much heavier
than the hyperon, it will have decay modes involving a vector meson;
this is certainly beyond the realm of current algebra. Because of
these two reasons, it is no longer justified to apply current algebra to
heavy-baryon weak decays, especially for $s$-wave amplitudes. Thus one has
to go back to the original pole model,
\foot{It is a ``model'' because of the assumption of pole approximation: The
nonfactorizable contribution is approximately saturated by  one-particle
intermediate states.}
which is nevertheless reduced to current algebra
in the soft pseudoscalar-meson limit, to deal with nonfactorizable
contributions. The merit of the pole model is obvious: Its use is very general
and is not limited to the soft meson limit and to the pseudoscalar-meson
final state. Of course, the price we have to pay is that it requires the
knowledge of the negative-parity baryon poles for the parity-violating
transition. This also explains why the theoretical study of nonleptonic
decays of heavy baryons is much more difficult than the hyperon and
heavy meson decays.

   Recently, a calculation of the nonfactorizable $s$- and $p$-wave
amplitudes of charmed baryon decays through the pole contributions from the
low-lying $\halfm$ resonances and ground-state $\halfp$ baryons has been
presented by us [14] and by Xu and Kamal [15]. We use the MIT bag model to
tackle both $\halfm$ and $\halfp$ baryon poles. By comparing the
pole-model and current-algebra results for the $s$ waves of $B_c\ri B+P$, we
reach an important conclusion: {\it the parity-violating amplitude of
charmed baryon decays is no longer dominated by the commutator terms.}
That is to say, away from the soft meson limit the correction to the
commutator terms is very important. This correction will affect the magnitude
and sometimes even the sign of the asymmetry parameter $\alpha$. Needless to
say, the pole model also allows us to treat the weak decays $B_c\ri B+V(1^-)$
on the same footing as $B_c\ri B+P(0^-)$ decays.

   In the previous publication [14] we have applied the pole model to some
selected decay modes, namely $\lamc\ri p\bar{K}^0(\bar{K}^{*0}),~\Lambda\pi^+(
\rho^+),~\Sigma^0\pi^+(\rho^+),~\Sigma^+\pi^0(\rho^0)$. The main purpose of
the present paper is to complete the pole-model analysis for all two-body
Cabibbo-allowed weak decays of the antitriplet charmed baryons $\lamc,~\xip,~
\xin$ and the sextet charmed baryon $\Omega_c^0$. Owing to large theoretical
uncertainties associated with the vector-meson case, as elaborated on
in detail in Ref.[14], we will confine ourselves to the decay modes $B_c\ri
B+P$.

   The present paper is organized as follows. The general framework of the
pole model is recapitulated in Section 2. Numerical results of the decay rate
and the asymmetry parameter $\alpha$ for Cabibbo allowed two-body nonleptonic
decays of charmed baryons are presented in Section 3 with some model details
given in Appendixes A-D. In Section 4 we then compare our results with current
algebra as well as recent theoretical calculation [15,16] and then draw
conclusions.

\chapter{General Considerations}
   Since the general framework for treating the nonleptonic weak decays of
charmed baryons is already discussed in Ref.[14], here we will  emphasize some
main points which are not thoroughly discussed in the previous publication.

   The QCD-corrected effective weak Hamiltonian responsible for the
Cabibbo-allowed charmed-baryon decays has the form
$$\H_{_W}=\,{G_F\over 2\sqrt{2}}V_{cs}V_{ud}(c_+O_++c_-O_-),\eqno(2.1)$$
with $O_{\pm}=(\bar{s}c)(\bar{u}d)\pm(\bar{s}d)(\bar{u}c)$, where $(\bar{q}_1
q_2)=\bar{q}_1\gamma_\mu(1-\gamma_5)q_2$, and $V_{ij}$ being the quark mixing
matrix element. The Wilson coefficients are evaluated at the charm mass scale
to be $c_+\cong 0.73$ and $c_-\cong 1.90$. In general the decay amplitude of
the baryon decay $B_i\ri B_f+P$ can be written in terms a sum
over intermediate hadronic states. As far as the vacuum intermediate state is
concerned, the amplitude will be factorized  if the pseudoscalar
meson $P$ can be created from the quark currents of $O_\pm$. (For a review of
factorization and the large $N_c$ approach, see e.g. Ref.[17]) Schemetically,
$$M(B_i\ri B_f+P)=\,M(B_i\ri  B_f+P)^\fac+M(B_i\ri B_f+P)^\nf,\eqno(2.2)$$
where the superscript n.f. stands for nonfactorization.
It is clear from the expression of $O_\pm$ that factorization occurs if the
final-state meson is the $\pi^+$ or $\bar{K}^0$. Explicitly,
$$\eqalign{
M(B_i\ri B_f+\pi^+)^\fac= &\,{G_F\over 2\sqrt{2}}V_{cs}V_{ud}\left(c_1+{c_2
\over N_c}\right)\bra{\pi^+}(\bar{u}d)\ket{0}\bra{B_f}(\bar{s}c)\ket{B_i}, \cr
M(B_i\ri B_f+\bar{K}^0)^\fac= &\,{G_F\over 2\sqrt{2}}V_{cs}V_{ud}\left(c_2+{c_1
\over N_c}\right)\bra{\bar{K}^0}(\bar{s}d)\ket{0}\bra{B_f}(\bar{u}c)\ket{B_i},
 \cr}\eqno(2.3)$$
where $N_c$ is the number of quark color degrees of freedom, and
$$c_1=\half(c_++c_-),~~~~c_2=\half(c_+-c_-).\eqno(2.4)$$
In the quark-diagram language, the $c_1$ ($c_2)$ term of the factorizable
$B_i\ri B_f+\pi^+$ ($B_i\ri B_f+\bar{K}^0$) amplitude comes from the
external (internal) $W$-emission diagram. The $W$-exchange diagram is of course
nonfactorizable.

  In the content of meson nonleptonic decay, it is customary to make a further
assumption, namely the factorization (or vacuum-saturation) approximation,
in which one only keeps the factorizable contributions and drops the
nonfactorizable ones. This approximation can be justified in the large $N_c$
limit [18] since the nonfactorizable amplitudes are suppressed, as far as the
color factor is concerned, by powers of $1/N_c$. However, in the $N_c\ri\infty$
limit, one should also drop the $1/N_c$-suppressed factorizable contribution
[see e.g. Eq.(2.3)]. The large $N_c$ version of factorization is thus
different from the naive factorization approximation in that the
Fierz-transformed
terms are taken into account in the latter approach. Nowadays we have
learned from the nonleptonic decays of charmed and bottom mesons that the naive
factorization method fails to account for the bulk of data, especially for
those decay modes which are naively expected to be color suppressed. The
discrepancy between theory and experiment gets much improved in the $1/N_c$
expansion method. Does this scenario also work for the baryon sector? This
issue is settled down by the experimental measurement of the Cabibbo-suppressed
mode $\lamc\ri p\phi$, which receives contributions only from the factorizable
diagrams. We have shown in Ref.[14] that the large-$N_c$ predicted rate is in
good agreement with the measured value. By contrast, its decay rate
prdicted by the naive factorization approximation is too small by a factor of
15. Therefore, we should take the $1/N_c$ approach for the factorizable
amplitude of $B_i\ri B_f+P$
$$\eqalign{ A^\fac=&\,-{G_F\over\sqrt{2}}V_{cs}V_{ud}f_{_P}c_k(m_{_{B_i}}
-m_f)f_1^{^{B_iB_f}}(m^2_{_P}),  \cr   B^\fac=&\,{G_F\over\sqrt{2}}V_{cs}V_{
ud}f_{_P}c_k(m_{_{B_i}}+m_f)g_1^{^{B_iB_f}}(m^2_{_P}),  \cr}\eqno(2.5)$$
where $f_{_P}$ is the decay constant of the meson $P$, $k=1$ for $\pi^+$
emission and $k=2$ for $\bar{K}^0$ emission, $f_1$ and $g_1$ are vector
and axial-vector form factors defined in Eq.(D1), and $A$ as
well $B$ are $s$- and $p$-wave amplitudes, respectively
$$M(B_i\ri B_f+P)=\,i\bar{u}_f(A-B\gamma_5)u_i.\eqno(2.6)$$

  We next turn to the nonfactorizable contribution. It is here we see a
significant disparity between meson and baryon decays. Contrary to the meson
case, the nonfactorizable amplitudes of baryon nonleptonic decays are not
necessarily color suppressed in the $N_c\ri\infty$ limit [16]. Although the
$W$-exchange diagram, for example, is down by a factor of $1/N_c$ relative to
the external $W$-emission diagram, this seeming suppression is compensated by
the fact that the baryon contains $N_c$ quarks in the limit of large $N_c$,
thus allowing $N_c$ different possibilities for $W$-exchange between heavy and
light quarks. This leads to the known
statement that $W$-exchange in baryon decay is subject to neither color nor
helicity suppression. Using the reduction formula, the nonfactorizable
amplitude can be recast to
$$ M(B_i\ri B_f+P^a(q))^\nf= \,\lim_{q^2\ri m_{_P}^2}i(m_{_P}^2-q^2)
\int d^4xe^{iq\cdot x}\bra{B_f}T\phi^a(x)\H_{_W}(0)\ket{B_i},\eqno(2.7)$$
or
$$\eqalign{ M(B_i\ri B_f+P^a(q))^\nf
=&\, \lim_{q^2\ri m_{_P}^2}i(m^2_{_P}-q^2)\int d^4xe^{iq\cdot x}\bigg(\sum_n
\theta(x^0)\bra{B_f}\phi^a(x)\ket{n} \cr \times &\bra{n}\H_{_W}(0)\ket{B_i}
+\sum_n\theta(-x^0)\bra{B_f}\H_{_W}(0)\ket{n}\bra{n}\phi^a(x)\ket{B_i}\bigg),
\cr}\eqno(2.8)$$
where $\phi^a$ is the interpolating field for the $P^a$. Conventionally one
considers pole approximation so that only one-baryon intermediate states
are kept.
Under the pole approximation, the nonfactorizable amplitude is nothing but
the contribution arising from two distinct pole diagrams. This can be seen from
the identity
$$\eqalign{ \bra{B_f}\phi^a(0)\ket{B_n}= &\,{1\over m_{_P}^2-q^2}\bra{B_f}J^a
(0)\ket{B_n},  \cr =&\,{ g_{_{B_fB_nP^a}}\over m_{_P}^2-q^2}\,\bar{u}_fi
\gamma_5u_n.  \cr}\eqno(2.9)$$
Hence, for example, the first term on the r.h.s. of (2.8) represents the pole
diagram
in which a weak transition $B_i-B_n$ is followed  by a strong emission of the
$P^a$. Note that since  the baryon-color wave function is totally
antisymmetric, only the operator $O_-$ contributes to the baryon-baryon
transition matrix element as it is antisymmetric in color indices. As shown in
Ref.[14], at least for hyperon and charmed-baryon decays, the $s$-wave
amplitude is dominated by the low-lying $\halfm$ resonances and the $p$-wave
one governed by the
ground-state $\halfp$ poles. As a result, it follows from Eq.(2.8) that [14]
$$\eqalign{A^\nf=&
-\sum_{B^*_n(\halfm)}\left({g_{_{B_fB_{n^*}
P}}b_{_{n^*i}}\over m_i-m_{n^*}}+{b_{_{fn^*}}g_{_{B_{_{n^*}}B_iP}}\over m_f-m_
{n^*}}\right)+\cdots,    \cr
 B^\nf=&-\sum_{B_n}\left({g_{_{B_fB_nP}}a_{_{ni}}\over m_i-m_n}+{a_{_{fn}}g_{_
{B_nB_iP}}\over m_f-m_n}\right)+\cdots,  \cr}\eqno(2.10)$$
where ellipses denote other pole contributions which are negligible for our
purposes, and $a_{ij}$ as well as $b_{i^*j}$ are the baryon-baryon matrix
elements defined by
$$\eqalign{ \bra{B_i}{\cal H}_{_W}\ket{B_j}= &\,\bar{u}_i(a_{ij}-b_{ij}
\gamma_5
)u_j,   \cr \bra{B^*_i(1/2^{^-})}{\cal H}^\pv_{_W}\ket{B_j} = &\,ib_{i^*j}\,
\bar{u}_iu_j, \cr }\eqno(2.11)$$
with $b_{ji^*}=-b_{i^*j}$. It should be stressed that Eq.(2.10) is derived only
under the assumption of pole approximation, and it is valid also for vector
meson emission.

  Evidently, the calculation of $s$-wave amplitudes is generally more
difficult than
the $p$-wave owing to the troublesome  negative-parity baryon resonances.
Nevertheless, a simplification happens for hyperon nonleptonic decays where
the final-state  pion is approximately soft. Using the Goldberger-Treiman (GT)
relation (C2) for the coupling constants $g_{_{BBP}}$ and the generalized GT
relation (C8) for $g_{_{B^*BP}}$ couplings (both relations being valid in
the soft pion limit), Eq.(2.8) leads to [14]
$$A^\ca=\,{\sqrt{2}\over f_{_{P^a}}}\bra{B_f}[Q_5^a,~\H_{_W}^\pv]\ket{B_i},
\eqno(2.12)$$
and
$$B^\ca=\,-{\sqrt{2}\over f_{_{P^a}}}\sum_{B_n}\left(g^A_{_{B_f
B_n}}\,{m_f+m_n\over m_i-m_n}\,a_{_{ni}}+a_{_{fn}}\,{m_i+m_n\over
m_f-m_n}\,g^A_{_{B_iB_n}}\right).\eqno(2.13)$$
Traditionally, the current-algebra results (2.12) and (2.13)
\foot{Eq.(2.7) together with (2.14) and (2.15) leads to
$$M(B_i\ri B_f+P^a)=-i{\sqrt{2}\over f_{_{P^a}}}\bra{B_f}[Q_5^a,~\H_{_W}]\ket
{B_i}+{\sqrt{2}\over f_{_{P^a}}}q^\mu T_\mu,$$
with $T_\mu=\int d^4xe^{iq\cdot x}\bra{B_f}TA_\mu^a(x)\H_{_W}(0)\ket{B_i}$.
Since $B^\nf$ and $q^\mu T_\mu$, the latter being the pole amplitude for
$B_i\ri B_f+A_\mu^a$,
are not well defined in the limit $q_\mu\ri 0$ but their difference does [19],
the current algebra expression for the parity-conserving wave should read
$$B^\ca=\lim_{q\ri 0}\left(B^\nf-i{\sqrt{2}\over f_{_{P^a}}}q^\mu T_\mu\right)+
i{\sqrt{2}\over f_{_{P^a}}}q^\mu T_{\mu},$$
which can be shown to be equivalent to Eq.(2.13). }
are derived from Eq.(2.8) together with the PCAC relation
$$\pi^a=\,{\sqrt{2}\over f_\pi m^2_{_P}}\partial^\mu A_\mu^a\eqno(2.14)$$
and the Ward identity
$$\eqalign{ i\int d^4xe^{iq\cdot x}T\partial^\mu A_\mu^a(x)\H_{_W}(0)= &\,
q^\mu\int d^4xTA_\mu^a(x)\H_{_W}(0)   \cr  -&\,\int d^4xe^{iq\cdot x}\delta
(x^0)[A_0^a(x),~\H_{_W}(0)].   \cr}\eqno(2.15)$$
Note that $B^\ca$ can actually be read off directly from Eq.(2.10) by
substituting the GT relation for strong coupling constants.
  Therefore, the parity-violating amplitude is reduced in the soft pion limit
to a simple commutator relation and is related to parity-conserving
baryon-baryon matrix elements. In other words, {\it no} information of $\halfm$
poles is required for evaluating the $s$-wave amplitudes. However, as
explained in Introduction, such simplicification is no longer applicable to
heavy-baryon weak decays for the meson there is far from being soft; for
example, the pion's momentum in the decay $\lamc\ri \Lambda\pi$ is 863 MeV,
which is much larger than its mass.
Writing
$$A=A^\ca+(A-A^\ca),\eqno(2.16)$$
it has been demonstrated in Refs.[14,15] that the on-shell correction $(A-A^
\ca)$ is very important for charmed-baryon decays, and this clearly
indicates that the $s$-wave amplitude is not dominated by the commutator term.

  To summarize, the dynamics of heavy-baryon decays is more complicated than
the meson decay because of the importance of nonfactorizable contributions,
and is more diffcult to treat than the hyperon decay owing to the presence
of $\halfm$ poles for $s$ waves. In short, Eq.(2.10) is the starting point
for handling the nonfactorization amplitudes of heavy baryon decays.

\chapter{Numerical Results}
    We employ the MIT bag model [20] to evaluate the form factors appearing
in factorizable amplitudes and the strong coupling constants and baryon
transition matrix elements relevant to nonfactorizable contributions. Some
model details are given in Appendixes A-D. In this section we will first
discuss the evaluation of the aforementioned ingredients and then present
the results of decay rates and the $\alpha$ asymmetry parameter for
Cabibbo-allowed nonleptonic weak decays of charmed baryons.
\section{Baryon-baryon transition matrix elements}

  Among the two four-quark operators $O_\pm$ given in Eq.(2.1), $O_+$ is
symmetric in color indices whereas $O_-$ is symmetric. Therefore, the operator
$O_+$ does not contribute to baryon transition matrix elements since the
baryon-color wave function is totally antisymmetric. The parity-conserving
(pc) matrix elements $a_{ij}$ and the parity-violating (pv) ones $b_{i^*j}$
have the expression
$$\eqalign{ a_{ij}=&\,{h\over 2\sqrt{2}}\bra{B_i}O_-^\pc\ket{B_j}c_-,  \cr
b_{i^*j}=&\,-i{h\over 2\sqrt{2}}\bra{B_i(1/2^-)}O_-^\pv\ket{B_j}c_-,  \cr}
\eqno(3.1)$$
with $h\equiv G_FV_{cs}V_{ud}$. Note that $b_{ji^*}=-b_{ij^*}$.
With the bag integrals $X_1=-3.58\times 10^{-6}~{\rm
GeV}^3$ and $X_2=1.74\times 10^{-4}~{\rm GeV}^3$ [14], the pc transitions
are (in units of $c_-h{\rm GeV}^3$)
$$\eqalign{ a_{_{\Sigma^+\lamc}}=&\,a_{_{\Sigma^0_c\Lambda^0}}=\,-3.76\times
10^{-3},~~~a_{_{\xin\Xi^0}}=\,-3.81\times 10^{-3},  \cr
a_{_{\Sigma^+_c\Sigma^+}}=&\,-a_{_{\Sigma^0_c\Sigma^0}}
=\,a_{_{\Xi_c^{0S}\Xi^0}}=\,-6.58\times 10^{-3},   \cr}\eqno(3.2)$$
where the superscripts $A$ and $S$ denote antitriplet and sextet charmed
baryons, respectively.

  In the bag model the low-lying negative-parity baryon states are made of two
quarks in the ground $1S_{1/2}$ eigenstate and one quark excited to $1P_{1/2}$
or $1P_{3/2}$. Consequently, the evaluation of the $\halfm-\halfp$ baryon
matrix elements $b_{i^*j}$ becomes much more involved owing to the presence of
$1P_{1/2}$ and $1P_{3/2}$ bag states. Assuming that the $\halfm$ resonances are
dominated by the low-lying negative-parity states, we have four ({\bf 70}, L=1)
states $^2\be_{1/2},~^4\be_{1/2},~^2\bten_{1/2},~^2{\bf 1}_{1/2}$ (see
Appendix A for notation) for uncharmed baryons and two states $^2\bsi_{1/2},~^2
{\bf \bar{3}}_{1/2}$ states for charmed bayons. With the bag
integrals given by Eq.(3.7) of Ref.[14], it follows from Eqs.(3.1), (A3),
(B2-B6), and Eq.(A7) of Ref.[14] that [Note that the pv matrix elements
$b_{_{\Sigma_c^*\Lambda}},~b_{_{\Sigma_c^*\Sigma}}$ presented in Ref.[14]
are for wrong SU(3) presentation (see also the footnote in Appendix A); they
are corrected here in Eq.(3.3).]
$$\eqalign{ b_{_{\Sigma^+(^28)\lamc}} = &\,-1.76\times 10^{-3},~~~b_{_{\Sigma
^+(^48)\lamc}}=\,-4.21\times 10^{-3},~~b_{_{\Sigma^+(^210)\lamc}}=\,1.47
\times 10^{-4},  \cr
b_{_{\Xi^0(^28)\xin}}=&\,4.77\times 10^{-5},~~~b_{_{\Xi^0(^48)\xin}}=-1.72
\times 10^{-3},~~~b_{_{\Xi^0(^210)\xin}}=-1.41\times 10^{-3},   \cr
b_{_{\Xi^0(^28)\xins}}=&\,3.55\times 10^{-3},~~~b_{_{\Xi^0(^48)\xins}}=7.02
\times 10^{-3},~~~b_{_{\Xi^0(^210)\xins}}=2.48\times 10^{-4},   \cr
b_{_{\Sigma^+(^26)\Lambda^0}}=&\,-7.97\times 10^{-4},~~~b_{_{\Sigma^0_c(^26)
\Sigma^0}}=\,1.46\times 10^{-3},~~~b_{_{\Sigma^+_c(^26)\Sigma^+}}
=\,-1.46\times 10^{-3},  \cr
b_{_{\Xi_c^0(^26)\Xi^0}}=&\,1.46\times 10^{-3},~~~b_{_{\Xi_c^0(^2\bar{3})\Xi^0
}}=-2.55\times 10^{-3},   \cr
b_{_{\Sigma_c^0(^26)\Sigma^0}}=&\,1.46\times 10^{-3},~~~b_{_{\Sigma_c^+(^2\bar
{3})\Sigma^+}}=-1.46\times 10^{-3},   \cr}\eqno(3.3)$$
expressed in units of $c_-h{\rm GeV}^3$.

\section{Form factors and strong coupling constants}

   Using the bag parameters given in Ref.[14], we obtain the following
values for the overlap bag integrals appeared in Eq.(D4)
$$\eqalign{
\int d^3r(u_su_c+v_sv_c)=&\,0.95,~~~~\int d^3r(u_uu_c+v_uv_c)=\,0.88,  \cr
\int d^3r(u_su_c-{1\over 3}v_sv_c)=&\,0.86,~~~~\int d^3r(u_uu_c-{1\over 3}v_u
v_c)=\,0.77\,.  \cr}\eqno(3.4)$$
The form factors $f_1$ and $g_1$ [see Eqs.(2.5) and (D1)] at $q^2=q^2_{\rm
max}=(m_i-m_f)^2$ then can be
determined directly from Eq.(D4) and  extrapolated to the desired $q^2$ using
Eq.(D3).

   In current algebra, strong coupling constants are related to the axial
vector form factors at $q^2=0$ via the Goldberger-Treiman (GT) relations given
by (C2) and (C8). With the bag integrals
$$ Z_1=\,0.052,~~~Z_2=\,0.056\,,\eqno(3.5)$$
the numerical values for the form factors $g_{B'B}^A$ and $g^A_{B_c'B_c}$ can
be read off immediately from Eqs.(C4) and (C6). Note that unlike the form
factor $g^A_{_{B_cBP}}$ (i.e. $g_1$), the $q^2$ dependence of $g^A_{_{B'B}}$
and $g^A_{_{B'_cB_cP}}$
is quite weak because of smallness of $q^2$. In what follows we list the
coupling constants $g_{_{B_c'B_cP}}$ calculated in this way:
$$\eqalign{
g_{_{\xip\xins\pi^+}}=&\,-14.3,~~~g_{_{\xin\xins\pi^0}}=\,10.3,~~~g_{_{\Sigma_c
^+\lamc\bar{K}^0}}=\,12.5,  \cr    g_{_{\lamc\Sigma_c^0\pi^+}}=&\,19.0,~~~
g_{_{\xip\Sigma^+_c\bar{K}^0}}=\,12.6,~~~g_{_{\xin\Sigma_c^+K^-}}=\,12.6,  \cr
g_{_{\lamc\xins K^+}}=&\,12.4,~~~g_{_{\xin\Sigma_c^0\bar{K}^0}}=\,17.8,~~~
g_{_{\xin\Omega_c^0\bar{K}^0}}=-18.7,   \cr  g_{_{\xins\Omega_c^0\bar{K}^0}}
=&\,22.5,~~~g_{_{\lamc\xin K^+}}=0.  \cr}\eqno(3.6)$$
The $g_{_{B'BP}}$ couplings computed by the method of Ref.[21] are summarized
in (C1). The reader may check that the current-algebra's predictions for
$g_{_{B'BP}}$ are smaller than those in (C1) by roughly a factor of
$\sqrt{2}$.

   The coupling constants $g_{_{B^*BP}}$ are obtained from
Eq.(C9) together with the generalized GT relation (C8). Taking the masses
\foot{The mass of $\Sigma(^28)$ and $\Sigma(^48)$ is taken from the Particle
Data Group [22]. In Ref.[14] we took $m_{_{\Sigma(^210)}}\simeq 2$ GeV, which
is unlikely to be the mass of the lowest-lying $(\bse,~^2\bten)$ state.}
$$\eqalign{
m_{_{\Sigma(^28)}}=&\,1620\,{\rm MeV},~~~m_{_{\Sigma(^48)}}=\,1750\,{\rm MeV},
{}~~~m_{_{\Sigma(^210)}}\simeq\,1700\,{\rm MeV},  \cr
m_{_{\Xi(^28)}}\simeq&\,1720\,{\rm MeV},~~~m_{_{\Xi(^48)}}\simeq\,1900\,{\rm
MeV},~~~m_{_{\Xi(^210)}}\simeq\,1800\,{\rm MeV},  \cr
m_{_{\Sigma_c(^26)}}\simeq &\,2750\,{\rm MeV},~~~m_{_{\Xi_c^*}}\simeq 2770\,
{\rm MeV}, \cr}\eqno(3.7)$$
for low-lying $\halfm$ resonances with $\Xi_c^*$ denoting $\Xi_c(^26)$ or
$\Xi_c(^2\bar{3})$, we obtain
$$\eqalign{
g_{_{\Sigma^+(^28)p\bar{K}^0}}=&\,0.52,~~~g_{_{\Sigma^+(^48)p\bar{K}^0}}=\,
2.49,~~~g_{_{\Sigma^+(^210)p\bar{K}^0}}=\,0.81,   \cr
g_{_{\Sigma^+(^28)\Xi^0K^+}}=&\,-0.47,~~~g_{_{\Sigma^+(^48)\Xi^0 K^+
}}=\,0.33,~~~g_{_{\Sigma^+(^210)\Xi^0 K^+}}=\,-0.15,   \cr
g_{_{\Sigma^+(^28)\Lambda^0\pi^+}}= &\,-0.63,~~g_{_{\Sigma^+(^48)\Lambda^0
\pi^+}}=\,1.59,~~g_{_{\Sigma^+(^210)\Lambda^0\pi^+}}=\,-1.11,  \cr
g_{_{\Sigma^+(^28)\Sigma^0\pi^+}}= &\,1.55,~~~~~g_{_{\Sigma^+(^48)\Sigma^0
\pi^+}}=\,0.81,~~g_{_{\Sigma^+(^210)p\bar{K}^0}}=\,-0.59,  \cr
g_{_{\Xi^0(^28)\Sigma^0\bar{K}^0}}=&\,-0.57,~~~g_{_{\Xi^0(^48)\Sigma^0\bar{K}
^0}}=\,0.39,~~~g_{_{\Xi^0(^210)\Sigma^0\bar{K}^0}}=\,-0.17,   \cr
g_{_{\Xi^0(^28)\Xi^-\pi^+}}=&\,0.41,~~~g_{_{\Xi^0(^48)\Xi^-\pi^+
}}=\,2.38,~~~g_{_{\Xi^0(^210)\Xi^-\pi^+}}=\,0.49,   \cr
g_{_{\Xi^0(^28)\Xi^0\pi^0}}=&\,0.29,~~~g_{_{\Xi^0(^48)\Xi^0\pi^0
}}=\,1.68,~~~g_{_{\Xi^0(^210)\Xi^0\pi^0}}=\,0.35,   \cr
g_{_{\Xi^0(^28)\Lambda\bar{K}^0}}=&\,0.57,~~~g_{_{\Xi^0(^48)\Lambda\bar{K}
^0}}=\,0.74,~~~g_{_{\Xi^0(^210)\Lambda\bar{K}^0}}=\,-0.32,   \cr}\eqno(3.8)$$
for couplings $g_{_{B'^*BP}}$, and
$$\eqalign{
g_{_{\Sigma_c^+(^26)\xin K^+}}=&\,0.13,~~~g_{_{\Sigma_c^+(^26)\xip \bar{K}^0}}
=0.13,~~~g_{_{\Sigma_c^0(^26)\xin\bar{K}^0}}=0.18,  \cr
g_{_{\xic(^26)\xip\pi^-}}=&\,-0.15,~~~g_{_{\xic(^2\bar{3})\xip\pi^-}}=\,
-0.26,~~~g_{_{\xic(^26)\xin\pi^0}}=\,0.11,  \cr
g_{_{\xic(^2\bar{3})\xin\pi^0}}=&\,0.18,~~~g_{_{\xic(^26)\lamc K^-}}=0.23,~~~
g_{_{\xic(^2\bar{3})\lamc K^-}}=-0.39,   \cr
g_{_{\Sigma_c^0(^26)\lamc\pi^-}}=&\,-0.72,~~~g_{_{\xic(^26)\Omega_c^0\bar{K}^0
}}=0.01,~~~g_{_{\xic(^2\bar{3})\Omega_c^0\bar{K}^0}}=-0.17,  \cr}\eqno(3.9)$$
for $g_{_{B_c'^*B_cP}}$ coupling constants,
where uses have been made of the bag integrals
$$\tilde{Y}_1=\,0.056,~~~\tilde{Y}'_1=\,0.058,~~~\tilde{Y}_{1s}=\,0.051,
\eqno(3.10)$$
and Eq.(A7) of Ref.[14].

\section{Decay rate and asymmetry parameter}
    Armed with all the necessary ingredients we are in position to compute the
pc and pv amplitudes from Eqs.(2.5) and (2.10) for all Cabibbo-allowed
nonleptonic decays $B_c\ri B+P$ with $B_c=\lamc,~\xin,~\xip,~\Omega_c^0$.
The decay rate and the up-down asymmetry parameter $\alpha$ are given by
$$\Gamma=\,{p\over 8\pi}\left\{{(m_i+m_f)^2-m^2_{_P}\over m^2_i}\,|A|^2+{
(m_i-m_f)^2-m^2_{_P}\over m^2_i}\,|B|^2\right\},\eqno(3.11)$$
with $p$ being the momentum of the meson in the rest frame of $B_i$, and
$$\alpha=\,{2\kappa{\rm Re}(A^*B)\over |A|^2+\kappa^2|B|^2}\eqno(3.12)$$
with $\kappa=p/(E_f+m_f)$. The calculated results are summarized in Table I.
In order
to have a feeling for the size of the branching ratio, we also calculate
this quantity using the lifetimes (except for $\Omega_c$)
$$\tau(\lamc)= 1.9\times 10^{-13}s,~~~\tau(\xin)=1.0\times 10^{
-13}s,~~~ \tau(\xip)= 4.1\times 10^{-13}s,\eqno(3.13)$$
where $\tau(\lamc)$ is taken from the Particle Data Group [22], $\tau(\xip)$
and $\tau(\xin)$ from the central values of recent E687 measurements [23].
\foot{The average value of $\tau(\xip)=(3.0^{+1.0}_{-0.6})\times 10^{-13}s$
cited by the Particle Data Group [22] is pulled low by the old NA-32 result
$\tau(\xip)=(0.20^{+0.11}_{-0.06})\,ps$.}
Experimental
results for the decay rates of $\lamc\ri p\bar{K}^0,~\Lambda\pi^+,~\Sigma^0
\pi^+,~\Xi^0 K^+$ (see Table III) are from Refs.[22,24].

   It is clear from Table III that the pole-model predictions are in good
agreement with experiment except for the decay $\lamc\ri\Xi^0 K^+$. In Sec.4.3
we will argue that presently we cannot make reliable predictions for the decay
modes $\lamc\ri\Xi^0 K^+$ and $\xin\ri\Sigma^+ K^-$. A detailed discussion of
our results and a comparsion with other works will be presented in Section 4.

\chapter{Discussion and Conclusion}
  Before drawing conclusions and implications from our predictions for charmed
baryon nonleptonic decays, it is pertinent to compare our results with the
traditional approach (pre-1992), namely current algebra, and the most
recent theoretical calculation (post-1992) presented in Refs.[15,16].

\section{Comparsion with current algebra}
   Except for Refs.[5,10] most previous studies on the dynamics of charmed
baryon two-body weak decays are based on the current-algebra technique. The
predictions are shown in Table II. The factorizable amplitudes
are the same as Table I. As for nonfactorizable contributions, the $s$-wave
amplitudes are calculated by using the commutator terms (E3-E4), while
the $p$-wave ones by Eq.(2.13).

    Although the current algebra/PCAC methods were widely employed before for
the study of $B_c\ri B+P$, several important improvements are made in the
present current-algebra calculation:
\item{(1)} As discussed in Sec.II, large-$N_c$ approximation rather than naive
factorization approximation, the former being supported by the experimental
measurement of $\lamc\ri p\phi$ decay, is utilized for describing the
factorizable
ampltiudes. This has an important consequence that the factorizable amplitude
of $B_c\ri B+\bar{K}^0$, which is naively expected to be color suppressed, is
no longer subject to color suppression and has a equal weight as
the factorizable amplitude of $B_c\ri B+\pi^+$. This helps explain why
the observed ratio of $\Gamma(\lamc\ri\Lambda\pi^+)/\Gamma(\lamc\ri p\bar{K}^0)
$ is smaller than unity.
\item{(2)} Form factors $f_1$ and $g_1$ evaluated by the static bag or quark
model, for example Eqs.(C3) and (D2), are interpretated as the predictions
obtained at maximum $q^2$ since static bag- or quark-model wave functions best
resemble the hadron state at $q^2=(m_i-m_f)^2$ where both baryons are static.
As a result, form factors at
$q^2=0$ become smaller than previously estimated.
The decay rate of $\lamc\ri\Lambda\pi^+$, which was
overestimated before by an order of magnitude or so (see Table III of
Ref.[14]), is now significantly reduced.
\item{(3)} Strong coupling constants and baryon matrix elements are calculated
using the bag model so that their relative signs are fixed. The relative signs
are important when different pole contributions are combined.
In many earlier publications, couplings and hadron matrix elements are often
related to each other through SU(3) symmetry. Somtimes this will result a
wrong relative
sign if care is not taken. A prominent example is the decay $\lamc\ri\Xi^0
K^+$, which receives very little contribution for its $s$ waves (see Table
II) and has the $p$-wave amplitude given by
$$\eqalign{ B^\ca(\lamc\ri\Xi^0K^+)
=&\,-{1\over f_{_K}}\bigg(g^A_{_{\Sigma^+\Xi^0}}{m_{_{\Sigma^+}}
+m_{_{\Xi^0}}\over m_{_\lamc}-m_{_{\Sigma^+}}}a_{_{\Sigma^+\lamc}}+a_{_{\Xi^0
\xin}}{m_{_\xin}+m_{_\lamc}\over m_{_{\Xi^0}}-m_{_\xin}}g^A_{_{\xin\lamc}}  \cr
&\,+a_{_{\Xi^0\xins}}{m_{_
\xins}+m_{_\lamc}\over m_{_{\Xi^0}}-m_{_\xins}}g^A_{_{\xins\lamc}}\bigg).
\cr}\eqno(4.1)$$
Since $g^A_{_{\xin\lamc}}=0$ [see Eq.(C7)], only the first and third terms in
(4.1) contribute to the current-algebra pc amplitude.
{}From Eqs.(3.2), (C4) and (C6) we find a large cancellation between these two
pole terms. By contrast, a large constructive interference was found
in Ref.[6] owing to wrong relative signs.
\item{(4)} The pc amplitude derived from the pole contribution of $i\sqrt{2}
q^\mu T_\mu/f_{_P}$ has the familiar expression [19]
$$B=-{\sqrt{2}\over f_{_P}}(m_i+m_f)\sum_{B_n}\left({g^A_{_{B_fB_n}}a_{ni}\over
m_i-m_n}+{a_{fn}g^A_{_{B_nB_i}}\over m_f-m_n}\right).\eqno(4.2)$$
As discussed in the footnote after Eq.(2.13), the contribution due to
$\lim_{q\ri 0}(B^\nf-i{\sqrt{2}\over f_{_P}}q^\mu T_\mu)$ should be taken
into account and it leads to Eq.(2.13) when combined with (4.2). This
correction is important for the decay modes $\lamc\ri \Sigma^0\pi^+,~\Xi^0
K^+$, and $\xin\ri \Sigma^0\bar{K}^0,~\Sigma^+ K^-,~\Lambda\bar{K}^0$.

   Recall that the predicted ratio of $\Gamma(\Lambda\pi^+)/\Gamma(p\bar{K}^0)$
in earlier attempts is considerably larger than unity, ranging from 2.3 to
13 (see Table III of Ref.[14]), while experimentally it is only $0.36\pm
0.20$ [21]. The improved current-algebra computation yields a value of 0.40
for this ratio
and a smaller absolute decay rate for $\lamc\ri\Lambda\pi^+$,
both being in the right ballpark.

   We now compare our work with current algebra. To compute the pc amplitudes
from Eq.(2.10) we actually apply the GT relation for the $g_{_{B'_cB_cP}}$
couplings and Eq.(C1) for the coupling constants $g_{_{B'BP}}$. The difference
between Tables I and II for the nonfactorizable $p$ waves thus comes from
the difference between $g_{_{B'BP}}$ and $\sqrt{2}(m_{_{B'}}+m_{_B})g^A_{_{
B'B}}/f_{_P}$. It is clear that pc amplitudes in Tables I and II are generally
the same except
for the channels $\lamc\ri\Sigma^0\pi^+,~\Sigma^+\pi^0,~\Xi^0K^+$ and $\xin
\ri\Sigma^+ K^-$. In both approaches, the nonfactorizable $W$-exchange
effects are not negligible; they are as important as the factorizable ones in
the decays $\xip\ri\Sigma^+\bar{K}^0,~\Xi^0\pi^+$, $\xin
\ri\Sigma^0\bar{K}^0$ and even dominate in the reactions $\xin\ri\Lambda
\bar{K}^0$ and $\Omega_c\ri\Xi^0\bar{K}^0$.

  The crucial difference between current algebra and the pole model lies in
the pv sector. By comparing Table I with  Table II, it is evident that
\item{(i)} the $s$-wave amplitudes are no longer dominated by the commutator
terms; that is, the on-shell correction $(A-A^\ca)$ is quite important and has
a sign opposite to that of $A^\ca$ [14,15],
\item{(ii)} the sign of the nonfactorizable pv amplitudes is opposite to
that predicted by current algebra for the decays $\lamc\ri p\bar{K}^0,~
\Sigma^0\pi^+,~\Sigma^+\pi^0$, indicating that $|A-A^\ca|>|A^\ca|$ in these
cases, and
\item{(iii)} for $\xip$ and $\xin$ decays, the commutator terms are of the same
equal weight as factorizable contributions, whereas nonfactorizable $s$ waves
are always suppressed in the pole model.

   The current-algebra method for $s$ waves is drastically simple as it does
not require the knowledge of excited $\halfm$ resonances. However, we see
that such a simplicification is certainly not applicable for describing the pv
 amplitudes of
charmed baryon weak decays as the pseudoscalar meson is no more soft.
We also see that the predicted signs of the total $s$-wave amplitudes of
$\lamc\ri\Sigma^0\pi^+,~\Sigma^+\pi^0$, $\xin\ri\Sigma^0\bar{K}^0$
and $\Omega_c^0\ri\Xi^0\bar{K}^0$ relative to the corresponding $p$ waves
are different in the pole model and
current algebra. Hence, even a measurement of the sign of the
$\alpha$ asymmetry parameter in above-mentioned decays would provide a very
useful test on various models. Experimentalists are thus urged to perform such
measurements.

\section{Comparsion with most recent theoretical calculation}
    There are two recent works [15,16] in which a complete analysis of $B_c\ri
B+P$ is performed and factorizable amplitudes are evaluated under the
large-$N_c$ approximation. Among these two works, the framework adopted by
Xu and Kamal (XK) [15] is most close to ours, while K\"orner and Kr\"amer (KK)
[16] chose to use the covariant quark model to tackle the three-body
transition amplitudes (instead of two-body transitions) directly. In this
subsection, a comparsion of our work with Refs.[15,16] will be made in order.

   Though XK employ the current-algebra's expression Eq.(4.2) to evaluate
the nonfactorizable $p$-wave amplitudes, they do consider the $\halfm$ pole
contributions to the $s$ waves. Their $s$-wave pole formula Eq.(14) is
identical to our Eq.(2.10) after applying the generalized GT relation (C8) for
the couplings $g_{_{B^*BP}}$ and $g_{_{B_c^*B_cP}}$.
XK used SU(3) and SU(4) symmetries to relate
the form factors $g^A_{_{B'BP}}$ and $g^A_{_{B'_cB_cP}}$ to the SU(3)
parameters $F$ and $D$, which are in turn determined from a fit to hyperon
semileptonic decays, and the diquark model to calculate the pc  baryon
matrix elements. It is the $s$-wave sector where the XK's work deviates mostly
from ours. XK argued that the product of form factors and pv matrix elements
for $\halfm-\halfp$ transitions can be related to pc baryon matrix elements.
Moreover, under the assumption that $(F^-+D^-)/(F^--D^-)\approx 0$, with
$F^-$ and $D^-$ being the analogues of the $F$ and $D$ parameters for
$\halfm-\halfp$ transition form
factors, they claimed that the $s$-wave pole contributions are completely
determined from the commutator terms and the masses of $\halfm$ resonances
without introducing further new parameters. In our analysis, we have applied
the MIT bag model to compute all the form factors and baryon-baryon matrix
elements involving $\halfm$ intermediate states.

  A comparsion of Table I with Tables I and II of Ref.[15] shows that we are
more or less in agreement  with XK on the $A^\pole$ amplitude in $\lamc$
decays except for $\lamc\ri p\bar{K}^0$, but our $A^\pole$ for $\xin,~\xip$
decays are dramatically different form those of XK not only in sign but also
in magnitude: ours being smaller by roughly an order of magnitude. It is not
clear to us what is the source of discrepancy.
\foot{A possibility is that the pv $\halfp-\halfp$ baryon matrix elements
$b_{ij}$ [cf. Eq.(2.11)] are important for $\xip,~\xin,~\Omega_c^0$ decays. It
has been shown
[8,9] that $b_{ij}$ are in general small for $\lamc\ri B+P$ decays, but they
have not yet been examined for other antitriplet charmed baryon decay.}
Since XK has larger $A^\pole$ for $\lamc\ri p\bar{K}^0$, which dominates over
$A^\fac$, their $\alpha$ is opposite to ours in sign (see Table III). Hence,
a measurement of the sign of $\alpha(\lamc\ri p\bar{K}^0)$ will furnish a
useful test on the importance of on-shell corrections to the $s$-wave
amplitude. Finally, we note that in spite of the disparity on the $\alpha$
parameter, the predicted decay rates by XK are nevertheless in accordance with
ours within a factor of 2.

   We next switch to the work of KK. Instead of decomposing the decay amplitude
into products of strong couplings and two-body weak transitions, KK analyze
the nonleptonic weak process using the spin-flavor structure of the
effective Hamiltonian and the wave functions of baryons and
mesons described by the covariant quark model.
The nonfactorizable amplitudes are then obtained
in terms of two wave function overlap parameters $H_2$ and $H_3$,
which are in turn determined by fitting to the experimental data of $\lamc\ri
p\bar{K}^0$ and $\lamc\ri\Lambda\pi^+$, respectively. Despite the absence of
first-principles calculation of the parameters $H_2$ and $H_3$, this quark
model approach has fruitful predictions for not only $B_c\ri B+P$, but also
$B_c\ri B+V,~B^*(\thalfp)+P$ and $B^*(\thalfp)+V$ decays.
Another advantage of this analysis is that each amplitude has one-to-one
quark-diagram interpretation.

It is clear from Table III that the predicted decay rates of $\xip\ri\Sigma^+
\bar{K}^0,~\xin\ri\Sigma^0\bar{K}^0$, $\Omega_c\ri\Xi^0\bar{K}^0$ by KK are
larger than ours and that of XK by an order of magnitude, whereas the decay
$\xin\ri\Xi^-\pi^+$ is strongly suppressed in the scheme of KK. Therefore, a
measurement of the ratios
$$R_1={\Gamma(\xip\ri\Sigma^+\bar{K}^0)\over \Gamma(\xip\ri\Xi^0\pi^+)},~~~
R_2={\Gamma(\xin\ri\Xi^0\pi^0)\over\Gamma(\xin\ri\Xi^-\pi^+)},~~~R_3={\Gamma(
\xin\ri\Sigma^0\bar{K}^0)\over\Gamma(\xin\ri\Xi^-\pi^+)},\eqno(4.3)$$
which are predicted to be respectively 0.21, 0.22, 0.11 in the pole model, and
1.83, 0.03, 1.13 in the covariant quark model, will be quite helpful to test
those two schemes.

\section{$\lamc\ri\Xi^0K^+$ and $\xin\ri\Sigma^+K^-$}
   The decays $\lamc\ri\Xi^0K^+$ and $\xin\ri\Sigma^+K^-$ share some common
features that they do not receive factorizable contributions and that their
$s$-wave amplitudes are very small and $p$-wave ones are subject to a large
cancellation. More explicitly,
$$\eqalign{ B^\pole(\lamc\ri\Xi^0K^+) =&\,-\left({g_{_{\Sigma^+\Xi^0K^+}}a_{_
{\Sigma^+\lamc}}\over m_{_\lamc}-m_{_{\Sigma^+}}}+{a_{_{\Xi^0
\xin}}g_{_{\lamc\xin K^+}}\over m_{_{\Xi^0}}-m_{_\xin}} +{a_{_{\Xi^0
\xins}}g_{_{\lamc\xins K^+}}\over m_{_{\Xi^0}}-m_{_\xins}}\right),  \cr
B^\pole(\xin\ri\Sigma^+K^-)= &\,-\left({g_{_{\Xi^0\Sigma^+K^-}}a_{_{\xin\Xi^0}}
\over m_{_{\xin}}-m_{_{\Xi^0}}}+{a_{_{\Sigma_c^+\Sigma^+}}g_{_{\xin\Sigma^+_c
K^-}}\over m_{_{\Sigma^+}}-m_{_{\Sigma_c^+}}}\right).  \cr}\eqno(4.4)$$
[The first line of Eq.(4.4) is identical to Eq.(4.1) after the use of the GT
relation.]
A substitution of Eqs.(C1), (3.2) and (3.6) into Eq.(4.4) clearly indicates
a large destructive interference in the $p$-wave amplitudes, resulting
rather small decay rates for both modes. The fact that the naive prediction
$\Gamma(\lamc\ri\Xi^0K^+)=1.1\times 10^9s^{-1}$ is too small compared to
the recent CLEO measurement [23] $(1.7\pm 0.4)\times 10^{10}s^{-1}$ shows that
our predictions for those two decays are unreliable. The situation becomes
even worse in the framework of current algebra (see Table II).
\foot{The discrepancy is improved in Ref.[15], but the prediction there is
still too small by a factor of 3 to 4 (see Table III). As noted in passing,
the $p$-wave formula used in Ref.[15] is that of Eq.(4.2).}

   The CLEO data thus suggest that the destructive interference in the $p$ wave
of $\lamc\ri\Xi^0K^+$ is not as severe as originally expected. There are
several
possibilities for allowing the alleviation of large cancellation. For example,
the Goldberger-Treiman relation for the coupling $g_{_{\lamc\xins K^+}}$ may
not work well, or the $g_{_{\lamc\xin K^+}}$ coupling constant is not strictly
zero, or the $\halfp$ resonances may make important contributions to the pc
amplitudes, or it requires the combination of above mechanisms. This issue
should be seriously concerned in the future study.

  It is worth mentioning that the predicted $\Gamma(\lamc\ri\Xi^0K^+)$ by KK
is in agreement with experiment. In the scheme of KK, the decay modes $\lamc
\ri\Xi^0K^+$ and $\xin\ri\Sigma^+K^-$ receive contributions only from the quark
diagrams IIa and III (see Ref.[16] for notation). KK observed that the
effect of diagram III is strongly suppressed relative to IIa. In other words,
these two decay modes proceed essentially through diagram IIa; strong
cancellation occurs only in diagram III.

   In the pole model, diagram IIa  corresponds to the pole diagram in which
a weak transition is followed by a strong emission of a meson, while diagram
III contributes to both different pole diagrams. Unfortunately, we do not know
how to separate diagram IIa from diagram III in the pole language. At any
rate, our goal is to understand the suppression  of diagram III in the pole
model in order to resolve the aforementioned problem.

\section{Conclusion}
   We now draw some conclusions from our analysis of nonleptonic weak
decays of charmed baryons into an octet baryon and a pseudoscalar meson.
\item{(i)} Large $N_c$ approximation for factorizable amplitudes, which works
well in the charmed- and bottom-meson sector, is also effective in the heavy
baryon
sector as borne out by the experimental measurement of $\lamc\ri p\phi$.
This accounts for the color nonsuppression of the decay $\lamc\ri p\bar{K}^0$
relative to $\lamc\ri\Lambda\pi^+$.
\item{(ii)} Nonfactorizable contributions are evaluated under pole
approximation so that they are saturated by one-particle intermediate states.
It turns out that $s$-wave amplitudes are dominated by the
excited $\halfm$ resonances, and $p$-wave ones by the ground-state $\halfp$
poles. In the soft pseudoscalar-meson limit, the parity-violating amplitude is
reduced to the current-algebra commutator term. We find that $s$ waves in
charmed baryon decays are no longer dominated by commutator terms; this is not
surprising since the meson is far from being soft. The important on-shell
correction $(A-A^\ca)$
will affect the $\alpha$ asymmetry parameter and changes its sign for the
decays $\lamc\ri\Sigma^0\pi^+,~\Sigma^+\pi^0$, $\xin\ri\Sigma^0\bar{K}^0$ and
$\Omega_c\ri\Xi^0\bar{K}^0$. Hence, even a measurment of the sign of $\alpha$
in these decay modes will discern current algebra and other theoretical
models.
\item{(iii)} Nonfactorizable $W$-exchange
effects are not negligible; they are comparable to the factorizable ones in
the decays $\xip\ri\Sigma^+\bar{K}^0,~\Xi^0\pi^+$, $\xin
\ri\Sigma^0\bar{K}^0$ and even dominate in the reactions $\xin\ri\Lambda
\bar{K}^0$ and $\Omega_c\ri\Xi^0\bar{K}^0$.
\item{(iv)} Form factors $f_1$ and $g_1$ evaluated by the staic bag or quark
model are interpretated as the predictions obtained at maximum $q^2$ where
both baryons are static. Consequently, form factors become smaller at $q^2=0$
than previously expected. The decay rate  of $\lamc\ri\Lambda\pi^+$, which was
largely overestimated before, is now significantly reduced.
\item{(v)} The decays $\lamc\ri\Xi^0K^+$ and $\xin\ri\Sigma^+K^-$ receive
dominant contributions from nonfactorizable $p$ waves. Owing to a large
cancellation in
the pole amplitude, we cannot make reliable predictions on their decay rates
and asymmetry parameters. An effort to resolve this problem is urgently needed.
\vskip 3.5cm

\centerline{\bf Acknowledgments}

\bigskip

One of us (H.Y.C.) would like to thank Prof. C. N. Yang and the
Institute for Theoretical Physics at Stony Brook for their hospitality
during his stay there where this work was finished. This research was
supported in part by the National Science Council of ROC under Contract No.
NSC82-0208-M001-001Y.

\endpage
\centerline{\bf Appendix A: Baryon Wave Functions}
\vskip 0.5 cm

  To fix the relative sign of the coupling constants, form factors,
parity-conserving and -violating matrix elements, it is very important to
employ the baryon wave functions consistently. In the present paper,
we use the isospin baryon-pseudoscalar coupling convention given in Ref.[25]
(see Appendix C) to fix the sign of the ground-state $\halfp$ octet baryon
wave functions.
In the following, we list those wave functions relevant to our purposes
$$\eqalign{ p=&\,{1\over \sqrt{3}}[\,uud\chi_s+(13)+(23)\,],  \cr
 \Sigma^{^+}=&\,-{1\over \sqrt{3}}[\,uus\chi_s+(13)+(23)\,],  \cr
 \Sigma^{^0}=&\,{1\over \sqrt{6}}[\,(uds+dus)\chi_s+(13)+(23)\,],  \cr
 \Lambda^{^0}=&\,-{1\over \sqrt{6}}[\,(uds-dus)\chi_{_A}+(13)+(23)\,],  \cr
 \Xi^{^0}=&\,{1\over\sqrt{3}}[\,ssu\chi_s+(13)+(23)\,],  \cr
 \Xi^-=&\,{1\over\sqrt{3}}[\,ssd\chi_s+(13)+(23)\,],  \cr
 \lamc=&\,-{1\over \sqrt{6}}[\,(udc-duc)\chi_{_A}+(13)+(23)\,],  \cr
 \Sigma^{^+}_c=&\,{1\over \sqrt{6}}[\,(udc+duc)\chi_s+(13)+(23)\,],  \cr
 \Sigma^{^0}_c=&\,{1\over \sqrt{3}}[\,ddc\chi_s+(13)+(23)\,],  \cr
 \Xi_c^{^{0A}}=&\,{1\over\sqrt{6}}[\,(dsc-sdc)\chi_{_A}+(13)+(23)\,],  \cr
 \Xi_c^{^{0S}}=&\,{1\over\sqrt{6}}[\,(dsc+sdc)\chi_s+(13)+(23)\,],  \cr
 \Xi_c^{^{+A}}=&\,{1\over\sqrt{6}}[\,(usc-suc)\chi_{_A}+(13)+(23)\,],  \cr
 \Xi_c^{^{+S}}=&\,{1\over\sqrt{6}}[\,(usc+suc)\chi_s+(13)+(23)\,],  \cr
 \Omega_c^{^0}=&\,{1\over\sqrt{3}}[\,ssc\chi_s+(13)+(23)\,],  \cr}\eqno(A1)$$
where $abc\chi_s=(2a^{\uparrow}b^\uparrow c^\downarrow-a^\up b^\dw c^\up-
a^\dw b^\up c^\up)/\sqrt{6},$ and $abc\chi_{_A}=(a^\up b^\dw c^\up-a^\dw b^\up
c^\up)/\sqrt{2}$, and the superscripts $A$ and $S$ indicate antitriplet and
sextet charmed baryons, respectively.

   The low-lying negative-parity $\halfm$ noncharmed baryons belong to the
(\bse, $L=1$) multiplet in the flavor-spin $SU(6)$ basis, which can be
decomposed into $SU(3)$ mutliplets as
$$\ket{\bse,~L=1}=\,\ket{\bse,~^2{\bf 8}_{1/2}}\oplus\ket{\bse,~^4{\bf 8}_{1/2}
}\oplus\ket{\bse,~^2{\bf 10}_{1/2}}\oplus\ket{\bse,~^2{\bf 1}_{1/2}},
\eqno(A2)$$
where the superscript and suberscript denote the quantum numbers $2S+1$ and
$J$, respectively.
In the MIT bag model these states are made of two quarks in the ground
$1S_{1/2}$ state and one quark excited to $1P_{1/2}$ or $1P_{3/2}$. That is,
the SU(6) (\bse, $L=1$) states can be constructed from the $\ket{{\bf 8},~P_{
1/2}}_a,~\ket{{\bf 8},~P_{1/2}}_b,~\ket{{\bf 8},P_{3/2}}$, $\ket{{\bf 10},P_{
1/2}}$ and $\ket{{\bf 10},P_{3/2}}$ configurations [26], where
$P_{1/2}\equiv(1S_{1/2})^21P_{1/2},~P_{3/2}\equiv(1S_{1/2})^21P_{3/2}$. The
explicit wave functions for the $\halfm$ resonances of $\Sigma^+$ are given by
Eq.(A8) of Ref.[14]. The wave functions for the low-lying negative-parity
states of the
octet baryons can be easily obtained from that of $\Sigma^+(\halfm)$ by an
appropriate replacement of quarks. For example, the $\Xi^0(\halfm)$ wave
functions may be obtained from $\Sigma^+(\halfm)$ wave functions by the
substitution $u\leftrightarrow s$.

   As for the charmed baryons, the charmed quark in the low-lying $\halfm$
state does not get excited, while the two light quarks of the charmed baryons
are either in the symmetric sextet or antisymmetric antitriplet state in the
SU(3) flavor space.
\foot{The SU(3) representation of charmed baryons given in Ref.[14] is
erroneous.}
The wave function of the $\halfm$ sextet charmed baryon, say $\Sigma^0_c(
\halfm)$, is simply given by
$$\ket{^2\bsi_{1/2},\halfm}=\,-{\sqrt{8}\over 3}\ket{\bsi,~P_{3/2}}-{1\over
3}\ket{\bsi,~P_{1/2}},\eqno(A3)$$
with
$$\eqalign{   \Sigma^0_c(\bsi,~P_{1/2})= &\,{1\over 6}\bigg\{2(\tddw\dup\cup+
\dup\tddw\cup)-\tdup\ddw\cup-\tdup\dup\cdw  \cr  & -\dup\tdup\cdw-\ddw\tdup\cup
+(13)+(23)\bigg\},  \cr      \Sigma^0_c(\bsi,~P_{3/2})= &\,{1\over 6}\bigg\{
\sqrt{3}(\ddw\bdupl\cdw+\bdupl\ddw\cdw)-\dup\bdup\cdw-\ddw\bdup\cup+\dup\bddw
\cup  \cr & -\bdup\dup\cdw-\bdup\ddw\cup+\bddw\dup\cup+(13)+(23)\bigg\},
\cr}\eqno(A4)$$
where the $1P_{1/2}~(1P_{3/2})$ quark is denoted by a tilde (undertilde), the
$s_z={3\over 2}$ quark state is remarked by $\bt{q}^{^{\big{\uparrow}}}$,
and $(ij)$ means permutation for the quark in place $i$ with the quark in place
$j$. The low-lying $\halfm$ resonance of the antitriplet charmed baryon, e.g.,
$\Xi_c^{^{0A}}(\halfm)$ has the form
$$\eqalign{\Xi_c^0(^2{\bf \bar{3}}_{1/2},\halfm)= &\,\Xi_c^0({\bf \bar{3}},
\halfm,P_
{1/2})  \cr  = &\,{1\over 2\sqrt{6}}\bigg\{ \dup\tsup\cdw-\ddw\tsup\cup+\tdup
\sdw\cup-\tdup\sup\cdw   \cr &-\sup\tdup\cdw+\sdw\tdup\cup-\tsup\ddw\cup
+\tsup\dup\cdw+(13)+(23)\bigg\}.  \cr}\eqno(A5)$$
The explicit spatial wave functions of the quark states $1S_{1/2},~1P_{1/2}$
and $1P_{3/2}$ are given in Appendix A of Ref.[14].

\vskip 1.0 cm
\centerline{\bf Appendix B: Parity-Conserving and -Violating Matrix Elements}
\vskip 0.5 cm

   Since the evaluation of the parity-conserving (pc) and parity-violating (pv)
matrix elements in the MIT bag model is already elaborated on in detail in
Appendix B
of Ref.[14], here we just summarize the matrix elements relevant to the
present paper. The pc matrix elements are found to be
\foot{Note that there is a sign misprint in Eq.(B4) of Ref.[14] which is
corrected here in Eq.(B1).}
$$\eqalign{ \bra{\Sigma^{^+}}O_-^\pc\ket{\lamc}=&\,\bra{\Lambda^0}O_-^\pc
\ket{\Sigma_c^0}=\,-{4\over\sqrt{6}}(X_1+3X_2)(4\pi),  \cr
\bra{\Sigma^{^+}}O_-^\pc\ket{\Sigma^{^+}_c}=&\,-\bra{\Sigma^{^0}}O_-^\pc
\ket{\Sigma^{^0}_c}=\,{2\sqrt{2}\over 3}(-X_1+9X_2)(4\pi), \cr
\bra{\Xi^{^0}}O_-^\pc\ket{\Xi_c^{^{0A}}}= &\,{4\over\sqrt{6}}(X_1-3X_2)(4\pi),
\cr \bra{\Xi^{^0}}O_-^\pc\ket{\Xi_c^{^{0S}}}= &\,-{4\over 3\sqrt{2}}(X_1+9X_2)
(4\pi),   \cr}\eqno(B1)$$
where $X_1$ and $X_2$ are the four-quark overlap bag integrals defined by
Eq.(B3) of Ref.[14].

   The evaluation of the parity-violating matrix elements for $\halfp-\halfm$
transitions is much more involved because the physical $\halfm$ baryon states
are linear combinations of $(S_{1/2})^2P_{1/2}$ and $(S_{1/2})^2P_{3/2}$ quark
eigenstates. Consequently, the number of the related bag overlap integrals is
largely increased. The relevant pv matrix elements for our purposes are
$$\eqalign{
\bra{\siga}O_-^\pv\ket{\lamc}=&\,i2\sqrt{2}\,(4\pi)(-{1\over 3}\tilde{X}_1+
\tx_2+{1\over 3}\tx_{1s}+\tx_{2s}),   \cr
\bra{\Sigma^{^+}(\be,P_{1/2})_b}O_-^\pv\ket{\lamc}=&\,i2\sqrt{2}
(4\pi)({2\over 3}\tx_1+{1\over 3}\tx_{1s}+\tx_{2s}),\cr
\bra{\sigc}O_-^\pv\ket{\lamc}=&\,-i{8\over 9}\sqrt{2\pi}(\bt{X}_{1}
+2\bt{X}_{1s}),   \cr
\bra{\Sigma^{^+}(\bten,P_{1/2})}O_-^\pv\ket{\lamc}=&\,i2\sqrt{2}
(4\pi)(-{1\over 3}\tx_1-\tx_2+{1\over 3}\tx_{1s}+\tx_{2s}),\cr
\bra{\Sigma^{^+}(\bten,P_{3/2})}O_-^\pv\ket{\lamc}=&\,i{8\over 9}\sqrt{4\pi}
(\bt{X}_1-\bt{X}_{1s}).  \cr}\eqno(B2)$$
for $\Sigma^{^+}(\halfm)-\lamc$ transitions,
$$\eqalign{
\bra{\Xi^{^0}(\be,P_{1/2})_a}O_-^\pv\ket{\Xi_c^{^{0A}}}=&\,i2\sqrt{2}(4\pi)
({1\over 3}\tx_1+\tx_2+{1\over 3}\tx_{1s}-\tx_{2s}),  \cr
\bra{\Xi^{^0}(\be,P_{1/2})_b}O_-^\pv\ket{\Xi_c^{^{0A}}}=&\,i2\sqrt{2}(4\pi)
({1\over 3}\tx_1+\tx_2-{2\over 3}\tx_{1s}),  \cr
\bra{\Xi^{^0}(\bten,P_{1/2})}O_-^\pv\ket{\Xi_c^{^{0A}}}=&\,i2\sqrt{2}(4\pi)
({1\over 3}\tx_1+\tx_2+{1\over 3}\tx_{1s}+\tx_{2s}),  \cr
\bra{\Xi^{^0}(\be,P_{3/2})}O_-^\pv\ket{\Xi_c^{^{0A}}}=&\,i{8\over 9}
\sqrt{2\pi}(2\bt{X}_1+\bt{X}_{1s}),   \cr
\bra{\Xi^{^0}(\bten,P_{3/2})}O_-^\pv\ket{\Xi_c^{^{0A}}}=&\,i{8\over 9}
\sqrt{4\pi}(\bt{X}_1-\bt{X}_{1s}),   \cr}\eqno(B3)$$
for $\Xi^{^0}(\halfm)-\Xi_c^{^{0A}}$ transitions,
$$\eqalign{
\bra{\Xi^{^0}(\be,P_{1/2})_a}O_-^\pv\ket{\Xi_c^{^{0S}}}=&\,i{4\over \sqrt{6}}
(4\pi)({1\over 3}\tx_1-3\tx_2-{1\over 3}\tx_{1s}-3\tx_{2s}),  \cr
\bra{\Xi^{^0}(\be,P_{1/2})_b}O_-^\pv\ket{\Xi_c^{^{0S}}}=&\,i{4\over \sqrt{6}}
(4\pi)({1\over 3}\tx_1-3\tx_2+{2\over 3}\tx_{1s}),  \cr
\bra{\Xi^{^0}(\bten,P_{1/2})}O_-^\pv\ket{\Xi_c^{^{0S}}}=&\,i{4\over \sqrt{6}}
(4\pi)({1\over 3}\tx_1-3\tx_2-{1\over 3}\tx_{1s}+3\tx_{2s}),  \cr
\bra{\Xi^{^0}(\be,P_{3/2})}O_-^\pv\ket{\Xi_c^{^{0S}}}=&\,i{8\over 9\sqrt{3}}
\sqrt{2\pi}(-2\bt{X}_1-\bt{X}_{1s}),   \cr
\bra{\Xi^{^0}(\bten,P_{3/2})}O_-^\pv\ket{\Xi_c^{^{0S}}}=&\,i{8\over 9\sqrt{3}}
\sqrt{4\pi}(-\bt{X}_1+\bt{X}_{1s}),   \cr}\eqno(B4)$$
for $\Xi^{^0}(\halfm)-\Xi_c^{^{0S}}$ transitions,
$$\eqalign{
\bra{\sigz_c(\bsi,P_{1/2})}O_-^\pv\ket{\Lambda^{^0}}= &\,i2\sqrt{3}(4\pi)
\,({1\over 3}\tx'_1+\tx'_2),\cr
\bra{\sigz_c(\bsi,P_{3/2})}O_-^\pv\ket{\Lambda^{^0}}= &\,i{4\sqrt{6}\over 9}
\sqrt{4\pi}(-\bt{X}'_1),  \cr
\bra{\sigz_c(\bsi,P_{1/2})}O_-^\pv\ket{\sigz}= &\,i2
(4\pi)\,({1\over 3}\tx'_1-3\tx'_2),  \cr
\bra{\sigz_c(\bsi,P_{3/2})}O_-^\pv\ket{\sigz}= &\,i{8\over 9}
\sqrt{2\pi}\,(-\bt{X}'_1),  \cr
\bra{\Sigma^{^+}_c(1/2^{^-})}O_-^\pv\ket{\Sigma^{^+}}=&\,-\bra{\sigz_c(
1/2^{^-})}O_-^\pv\ket{\sigz},  \cr}\eqno(B5)$$
for $\Sigma_c^0(\halfm)-\Lambda^0$ and $\Sigma_c(\halfm)-\Sigma$ transitions,
and
$$\eqalign{
\bra{\Xi_c^{^0}(\bsi,P_{1/2})}O_-^\pv\ket{\Xi^{^0}}= &\,i2(4\pi)({1\over 3}
\tx'_1-3\tx'_2),  \cr
\bra{\Xi_c^{^0}(\bsi,P_{3/2})}O_-^\pv\ket{\Xi^{^0}}= &\,i{8\over 9}\sqrt{2\pi}
(-\bt{X}'_1),  \cr
\bra{\Xi_c^{^0}({\bf \bar{3}},P_{1/2})}O_-^\pv\ket{\Xi^{^0}}= &\,i2\sqrt{3}
(4\pi)({1\over 3}\bt{X}'_1-\bt{X}'_2),  \cr}\eqno(B6)$$
for $\Xi_c^0(\halfm)-\Xi^0$ transitions, where the bag integrals $\tx_1,~\tx_2,
{}~\tx_{1s},~\tx_{2s},~\bt{X}_1,~\bt{X}_{1s}$, $\tx'_i,~\bt{X}'_i$  are defined
in Appendix B of Ref.[14].

\vskip 0.7 cm
\centerline{\bf Appendix C: Strong Coupling Constants}
\vskip 0.5 cm

   The octet baryon-pseudoscalar meson $BBP$ coupling constants can be reliably
evaluated using the method of Ref.[21] which employes the null result of
Coleman and Glashow for the tadpole-type symmetry breaking. The results
related to the present paper are
$$\eqalign{ g_{_{\Sigma^+p\bar{K}^0}}= &\,4.9,~~~g_{_{\Sigma^+\Lambda^0\pi^+}}
=\,11.8,~~~g_{_{\Sigma^+\Xi^0K^+}}=\,25.6,  \cr
g_{_{\Xi^0\Sigma^0\bar{K}^0}}=&\,-18.1,~~~g_{_{\Xi^0\Xi^-\pi^+}}=\,-6.1,~~~
g_{_{\Xi^0\Lambda\bar{K}^0}}=\,5.6,  \cr
g_{_{\Xi^0\Xi^0\pi^0}}=&\,-4.3,~~~g_{_{\Sigma^+\Sigma^+\pi^0}}
=\,-g_{_{\Sigma^+\Sigma^0\pi^+}}=\,13.3,  \cr}\eqno(C1)$$
where the sign of the coupling constants is fixed by the isospin coupling
convention given in Ref.[25]. As shown in Ref.[21], the above $g_{_{B'BP}}$
couplings are in good agreement with experiment.
The quantity of interest in the approach of current algebra is
$g^A_{_{B'B}}$, the axial-vector form factor at $q^2=0$, which is related to
the $BBP$ coupling constant via the Goldberger-Treiman (GT) relation
$$g_{_{B'BP^a}}=\,\sqrt{2}\,{m_{_{B'}}+m_{_B}\over f_{_{P^a}}}\,g^A_{_{B'B}},
\eqno(C2)$$
where $f_{_{P^a}}$ is the decay constant of the pseudoscalar meson $P^a$ (
$a=1,\cdots 8$) in the SU(3) representation. Note that the axial-vector
current corresponding to, for example $P^3$, is $\half(\bar{u}\gamma_\mu
\gamma_5 u-\bar{d}\gamma_\mu\gamma_5d)$.
In the bag model the axial form factor in static limit is given by
$$g_{_{B'B}}^A=\,\bra{B'\up}b^\dagger_{q_1}b_{q_2}\sigma_z\ket{B\up}\int d^3r
(u_{q_1}u_{q_2}-{1\over 3}v_{q_1}v_{q_2}),\eqno(C3)$$
where $u(r)$ and $v(r)$ are respectively the large and small components of
the wave function for the quark state $1P_{1/2}$.
We find
$$\eqalign{ g^A_{_{\Sigma^+p}}=& {1\over 5}g^A_{_{\Sigma^+\Xi^0}}=-{\sqrt{2}
\over
 5}g^A_{_{\Sigma^0\Xi^0}}={\sqrt{2\over 3}}g^A_{_{\Xi^0\Lambda^0}}={4\pi\over
3}Z_1,  \cr   g^A_{_{\Xi^0\Xi^-}}= &\, 2g^A_{_{\Xi^0\Xi^0}}=-{1\over\sqrt{6}}
g^A_{_{\Sigma^+\Lambda^0}}={1\over 2\sqrt{2}}g^A_{_{\Sigma^+\Sigma^0}}
= -{1\over 2}g^A_{_{\Sigma^+\Sigma^+}}=-{4\pi\over 3}Z_2,
\cr}\eqno(C4)$$
where
$$\eqalign{Z_1= &\,\int r^2dr(u_u^2-{1\over 3}v_u^2),  \cr
Z_2=&\,\int r^2dr(u_uu_s-{1\over 3}v_uv_s).  \cr}\eqno(C5)$$

   As for charmed baryon-pseudoscalar $B_cB_cP$ coupling, we will rely on
the GT relation (C2). The results are
$$\eqalign{  g^A_{_{\Xi^+_c\Xi_c^{0s}}}= &\,-2g^A_{_{\Xi_c^{0A}\Xi_c^{0s}}}=
\,-{1\over \sqrt{2}}g^A_{_{\Sigma_c^+\lamc}}=\,-{1\over \sqrt{2}}g^A_{_{
\lamc\Sigma_c^0}}=\,-{4\pi\over\sqrt{3}}Z_1,   \cr
g^A_{_{\Xi_c^+\Sigma_c^+}}=&\,-g^A_{_{\Xi_c^{0A}\Sigma_c^+}}=\,-g^A_{_{\Xi_c
^{0s}\lamc}}=\,{1\over\sqrt{2}}g^A_{_{\Xi_c^{0A}\Omega_c^0}}  \cr
=&\,-{1\over\sqrt{2}}g^A_{_{\Xi_c^{0A}\Sigma_c^0}}=\,-{\sqrt{3}\over 2\sqrt{2}}
g^A_{_{\Xi_c^{0s}\Omega_c^0}}=\,-{4\pi\over\sqrt{3}}Z_2,  \cr}\eqno(C6)$$
and
$$g^A_{_{B_{\bar{3}}B_{\bar{3}}}}=0,~~~~~{\rm for}~B_{\bar{3}}=\lamc,~\xin,~
\xip.\eqno(C7)$$
Interestingly, Eq.(C7) is a rigorous and model-independent statement in the
infinite charmed-quark mass limit. This comes from the fact that the light
diquark in the ${\bf \bar{3}}$ multiplet has spin parity $0^+$ and that the
pseudoscalar meson is emitted solely from the light quarks in the heavy quark
limit.
Since the transition $0^+\ri 0^++P$ does not conserve parity, it leads to
vanishing $B_{\bar{3}}B_{\bar{3}}P$ coupling.

  To evaluate the $s$-wave amplitudes we also need to know the $B^*BP$
coupling constants ($B^*$: $\halfm$ resonance). We shall use the
generalized GT relation
$$g_{_{B^*BP^a}}=\,\sqrt{2}\,{m_{_{B^*}}-m_{_B}\over f_{_{P^a}}}\,g^A_{_{B^*B}
},\eqno(C8)$$
to estimate the couplings $g_{_{B^*BP^a}}$.
It has been shown that this generalized GT relation, when applied
to the $\Lambda^{^*}\Sigma^{^+}\pi^{^+}$ interaction, is in good agreement with
experiment [27]. Note that $g_{_{BB^*P}}=g_{_{B^*BP}}$, while
$g^A_{_{BB^*}}=-g^A_{_{B^*B}}$. In the static limit, we find
$$\eqalign{ g^A_{_{B^*B}}= &\,\int r^2dr(\tilde{v}v-\tilde{u}u)\int d\Omega
\bra{B^*}b^\dagger_{\tilde{q}}b_q\ket{B}\cr  +&\,\int r^2dr(\bt{v}v-\bt{w}u)
\int d\Omega\bra{B^*}b^\dagger_{\bt{q}}b_q(\sigma_z
\hat{r}_z)\ket{B}.  \cr}\eqno(C9)$$
Since $\bra{B^*(P_{3/2})}b^\dagger_{\bt{q}}b_q(\sigma\hat{r}_z)\ket{B(S_{1/2})
}=0$, it is clear that $g^A_{_{B^*B}}$ is determined by the matrix element
$\int d\Omega\bra{B^*}b^\dagger_{\tilde{q}}b_q\ket{B}$ and the overlap
integrals
$$\eqalign{ \tilde{Y}_1=&\,\int r^2dr(\tilde{u}_uu_u-\tilde{v}_uv_u),  \cr
 \tilde{Y}_1'=&\,\int r^2dr(\tilde{u}_uu_s-\tilde{v}_uv_s),  \cr
 \tilde{Y}_{1s}=&\,\int r^2dr(\tilde{u}_su_u-\tilde{v}_sv_u).  \cr}
\eqno(C10)$$

\vskip 0.7 cm
\centerline{\bf Appendix D: Form Factors}
\vskip 0.4cm

  To evaluate the factorizable amplitudes of baryon weak decays requires the
information on the form factors $f_1$ and $g_1$ defined by
$$\eqalign{\bra{B_f}V_\mu-A_\mu\ket{B_i}=&\,\bar{u}_f(p_f)[f_1\gamma_\mu+
if_2\sigma_{\mu\nu}q^\nu+f_3q_\mu   \cr   &-g_1\gamma_\mu\gamma_5-ig_2\sigma_
{\mu\nu}q^\nu\gamma_5-g_3q_\mu\gamma_5]u_i(p_i),  \cr}\eqno(D1)$$
with $q_\mu=(p_i-p_f)_\mu$. In the static limit $f_1$ and $g_1$ are derived
in the bag model to be
$$f_1^{^{B_fB_i}}=\,\bra{B_f\up}b_{q_1}^\dagger b_{q_2}\ket{B_i\up}\int d^3r(
u_{q_1}u_{q_2}+v_{q_1}v_{q_2})\eqno(D2)$$
and Eq.(C3). However, contrary to the conventional interpretation, (D2) and
(C3) should be regarded as the bag-model predictions obtained at maximum
four-momentum transfer squared, i.e. $q^2=(m_i-m_f)^2$. This is because
the static-bag wave functions best resemble hadronic states in
the frame where both baryons are {\it static}. This can be achieved by
choosing the Breit frame where ${\bf p}_i={\bf p}_f={\bf q}/2=0$.

   For definiteness, we will assume a dipole $q^2$ dependence for the form
factors
$$f_1(q^2)=\,{f_1(0)\over (1-q^2/m^2_{_V})^2},~~~~g_1(q^2)=\,{g_1(0)\over (1-
q^2/m^2_{_A})^2},\eqno(D3)$$
where the pole masses are $m_{_V}(1^-)=2.01$ GeV, $m_{_A}(1^+)=2,42$ GeV for
the pole with the quark content $(c\bar{d})$ [22] and $m_{_V}(1^-)=2.11$ GeV
and $m_{_A}(1^+)=2.54$ GeV for the pole with the $(c\bar{s})$ quark content.
We find at $q^2=q^2_{\rm max}=(m_i-m_f)^2$ that
$$\eqalign{
f_1^{^{\lamc\Lambda}}=&\,\sqrt{2\over 3}f_1^{^{\xin\Xi^-}}=-\sqrt{2\over
3}f_1^{^{\xip\Xi^0}}=\int d^3r(u_su_c+v_sv_c),  \cr
f_1^{^{\Omega_c^0\Xi^0}}=&\,-\sqrt{2\over 3}f_1^{^{\lamc p}}=-\sqrt{2\over 3}
f_1^{^{\xip\Sigma^+}}={2\over\sqrt{3}}f_1^{^{\xin\Sigma^0}}=2f_1^{^{\xin
\Lambda}}=\int d^3r(u_uu_c+v_uv_c),  \cr
g_1^{^{\lamc\Lambda}}=&\,\sqrt{2\over 3}g_1^{^{\xin\Xi^-}}=-\sqrt{2\over
3}g_1^{^{\xip\Xi^0}}=\int d^3r(u_su_c-{1\over 3}v_sv_c),  \cr
-3g_1^{^{\Omega_c^0\Xi^0}}=&\,-\sqrt{2\over 3}g_1^{^{\lamc p}}=-\sqrt{2\over 3}
g_1^{^{\xip\Sigma^+}}={2\over\sqrt{3}}g_1^{^{\xin\Sigma^0}}=2g_1^{^{\xin
\Lambda}}=\int d^3r(u_uu_c-{1\over 3}v_uv_c).  \cr}\eqno(D4)$$
With the overlap bag integrals given by Eq.(3.4) it is straightforward to
check that our numerical results for form factors extrapolated to $q^2=0$
are in agreement with Table VI of
Ref.[28] for $\lamc\ri\Lambda,~\xip\ri\Xi^0$ and $\xin\ri\Xi^-$ transitions.
Form factors induced by the $c\ri u$ current are not gievn in Ref.[28]. If
(D2) and (C3) were interpretated as bag predictions at $q^2=0$, the calculated
branching ratio of the exclusive $\lamc\ri\Lambda$ decay would have been
enhanced by a factor of 3.5, which is in violent disagreement with experiment
[14]. This is another indication that the static-bag calculation of form
factors is indeed carried out at maximum $q^2$ rather than at $q^2=0$.

\vskip 0.7cm
\centerline{\bf Appendix E: Current Algebra Commutator Terms}
\vskip 0.4cm

  In current algbra the nonfactorizable $s$-wave amplitude of the decay
$B_c\ri B+P^a$ in
the soft meson limit is governed by the commutator term
$$\aca=\,-{\sqrt{2}\over f_{_{P^a}}}\bra{B}[Q^a_5,~\H^\pv]\ket{B_c},\eqno(E1)$$
where $f_\pi=132$ MeV, and $f_{_K}=1.22f_\pi$. As an example, consider the
decay $\xin\ri\Lambda\bar{K}^0$,
$$\aca(\xin\ri\Lambda\bar{K}^0)=\,-{1\over f_{_K}}\bra{\Lambda}Q^{K^0}\H^\pc
\ket{\xin}.\eqno(E2)$$
{}From Eq.(A1) we obtain $\bra{\Lambda}Q^{K^0}=\sqrt{3\over 2}\bra{\Xi^0}$ and
hence
$$\aca(\xin\ri\Lambda\bar{K}^0)=\,-\sqrt{3\over 2}\,{1\over f_{_K}}\bra{\Xi^0}
\H^\pc\ket{\xin}.\eqno(E3)$$

   The remaining $s$-wave commutator terms are summarized below:
$$\eqalign{\aca(\lamc\ri p\bar{K}^0)=&\,{1\over f_{_K}}\bra{\Sigma^+}\H^\pc
\ket{\lamc},  \cr   \aca(\lamc\ri\Lambda\pi^+)= &\,0,  \cr
\aca(\lamc\ri\Sigma^0\pi^+)=&\,-\aca(\lamc\ri\Sigma^+\pi^0)=\,{\sqrt{2}\over
f_\pi}\bra{\Sigma^+}\H^\pc\ket{\lamc},  \cr
\aca(\lamc\ri\Xi^0K^+)=&\,-{1\over f_{_K}}\bra{\Sigma^+}\H^\pc\ket{\lamc}+
{1\over f_{_K}}\bra{\Xi^0}\H^\pc\ket{\xin},  \cr
\aca(\xip\ri\Xi^0\pi^+)=&\,{1\over f_\pi}\bra{\Xi^0}\H^\pc\ket{\xin},  \cr
\aca(\xip\ri\Sigma^+\bar{K}^0)=&\,-{1\over f_{_K}}\bra{\Sigma^+}\H^\pc\ket{
\lamc},  \cr
\aca(\xin\ri\Sigma^0\bar{K}^0)=&\,{1\over \sqrt{2}f_{_K}}\bra{\Xi^0}\H^\pc\ket
{\xin},  \cr
\aca(\xin\ri\Sigma^+K^-)=&\,-{1\over f_{_K}}\bra{\Xi^0}\H^\pc\ket{\xin}+{1\over
f_{_K}}\bra{\Sigma^+}\H^\pc\ket{\lamc},  \cr
\aca(\xin\ri\Xi^0\pi^0)=&\,-{\sqrt{2}\over f_\pi}\bra{\Xi^0}\H^\pc\ket{\xin},
\cr
\aca(\xin\ri\Xi^-\pi^+)=&\,-{1\over f_\pi}\bra{\Xi^0}\H^\pc\ket{\xin},  \cr
\aca(\Omega_c^0\ri\Xi^0\bar{K}^0)=&\,{\sqrt{2}\over f_{_K}}\bra{\Xi^0}\H^\pc
\ket{\Xi_c^{^{0S}}}.  \cr}\eqno(E4)$$

\endpage

\ref{For a review of charmed baryons, see J.G. K\"orner and H.W. Siebert,
{\sl Ann. Rev. Nucl. Part. Sci.} {\bf 41}, 511 (1991);
S.R. Klein, {\sl Int. J. Mod. Phys.} {\bf A5}, 1457 (1990).}
\ref{J.N. Butler, invited talk presented at the XXVI International Conference
on High Energy Physics, Dallas, August 1992.}
\ref{L.L. Chau, {\sl Phys. Rep.} {\bf 95}, 1 (1983); L.L. Chau and H.Y. Cheng,
\prl {\bf 56}, 1655 (1986); \pr {\bf D36}, 137 (1987).}
\ref{L.L Chau, H.Y. Cheng, and B. Tseng, IP-ASTP-04-92.}
\ref{J.G. K\"orner, G. Kramer, and J. Willrodt, \pl {\bf 78B}, 492 (1978);
\zp {\bf C2}, 117 (1979).}
\ref{B. Guberina, D. Tadi\'c, and J. Trampeti\'c, \zp {\bf C13}, 251 (1982).}
\ref{F. Hussain and M.D. Scadron, {\sl Nuovo Cimento} {\bf 79A}, 248 (1984);
F. Hussain and K. Khan, {\sl ibid} {\bf 88A}, 213 (1985); R.E. Karlsen and
M.D. Scadron, {\sl Europhys. Lett.} {\bf 14}, 319 (1991).}
\ref{D. Ebert and W. Kallies,  {\sl Phys. Lett.} {\bf 131B}, 183 (1983);
{\bf 148B}, 502(E) (1984); {\sl Yad. Fiz.} {\bf 40}, 1250 (1984);
{\sl Z. Phys.} {\bf C29}, 643 (1985).}
\ref{H.Y. Cheng, \zp {\bf C29}, 453 (1985).}
\ref{Yu.L. Kalinovsky, V.N. Pervushin, G.G. Takhtamyshev, and N.A. Sarikov,
{\sl  Sov. J. Part. Nucl.} {\bf 19}, 47 (1988), and references therein.}
\ref{S. Pakvasa, S.F. Tuan, and S.P. Rosen, \pr {\bf D42}, 3746 (1990).}
\ref{G. Kaur and M.P. Khanna, \pr {\bf D44}, 182 (1991); {\sl ibid} {\bf D45},
R3024 (1992).}
\ref{G. Turan and J.O. Eeg, \zp {\bf C51}, 599 (1991).}
\ref{H.Y. Cheng and B. Tseng, \pr {\bf D46}, 1042 (1992).}
\ref{Q.P. Xu and A.N. Kamal, \pr {\bf D46}, 270 (1992).}
\ref{J.G. K\"orner and M. Kr\"amer, \zp {\bf C55}, 659 (1992).}
\ref{H.Y. Cheng, {\sl Int. J. Mod. Phys.} {\bf A4}, 495 (1989).}
\ref{ A.J. Buras, J.-M. G\'erard, and R. R\"uckl, \np {\bf B268}, 16 (1986).}
\ref{See, e.g., D. Bailin, {\it Weak Interactions} (Sussex University Press,
1977).}
\ref{A. Chodos, R.L. Jaffe, K. Johnson, and C.B. Thorn, \pr {\bf D10}, 2599
(1974); T. DeGrand, R.L. Jaffe, K. Johnson, and J. Kiskis, {\sl ibid} {\bf
D12}, 2060 (1975).}
\ref{M.P. Khanna and R.C. Verma, \zp {\bf C47}, 275 (1990).}
\ref{Particle Data Group, \pr {\bf D45}, S1 (1992).}
\ref{P.L. Frabetti {\it et al}., \prl {\bf 70}, 1381 (1993); {\sl ibid} {\bf
70}, 2058 (1993).}
\ref{V. Jain, VUHEP 92-02 (1992).}
\ref{G. Campbell, Jr, \pr {\bf D13}, 662 (1976).}
\ref{T.A. DeGrand and R.L. Jaffe, {\sl Ann. Phys.} (N. Y.) {\bf 100}, 425
(1976); T.A. DeGrand, {\sl ibid} {\bf 101}, 496 (1976).}
\ref{S. Pakvasa and J. Trampeti\'c, \pl {\bf 126B}, 122 (1983).}
\ref{R. P\'erez-Marcial, R. Huerta, A. Garc\'ia, and M. Avila-Aoki, \pr {\bf
D40}, 2955 (1990); {\sl ibid} {\bf D44}, 2203(E) (1991).}
\endpage
\refout
\endpage
\centerline{\bf Table Captions}
\vskip 1.5 cm

\item{{\rm Tab.~1}.} Numerical values of the predicted $s$- and $p$-wave
amplitudes of
$B_c\rightarrow B+P$ decays in the pole model in units of $G_FV_{cs}V_{ud}
\times 10^{-2}{\rm GeV}^2$. The predicted $\alpha$ asymmetry parameter,
decay rates (in units of $10^{11}s^{-1}$) and branching ratios (in percent)
are given in the last three columns. Lifetimes of charmed baryons are taken
from Eq.(3.13). As discussed in Sec.4.3, no reliable predictions can be made
for the decays $\lamc\to\Xi^0K^+$ and $\xin\to\Sigma^+K^-$.
\item{{\rm Tab.~2}.} Same as Table I except that predictions are made by
current algebra.
\item{{\rm Tab.~3}.} The predicted decay rates (in units of $10^{11}s^{-1}$)
and the $\alpha$ asymmetry parameter (in parentheses) for $B_c\rightarrow B+P$
decays in various models.

\end

%
%

\def\pole{{\rm pole}}
\def\com{{\rm com}}
\def\fac{{\rm fac}}
\def\tot{{\rm tot}}
\def\lamc{\Lambda^+_c}
\def\xia{\Xi^{+A}_c}
\def\xi0a{\Xi^{0A}_c}

\hsize=9 in

Table I. Numerical values of the predicted $s$- and $p$-wave amplitudes of
$B_c\rightarrow B+P$ decays in the pole model in units of $G_FV_{cs}V_{ud}
\times 10^{-2}{\rm GeV}^2$. The predicted $\alpha$ asymmetry parameter,
decay rates (in units of $10^{11}s^{-1}$) and branching ratios (in percent)
are given in the last three columns. Lifetimes of charmed baryons are taken
from Eq.(3.13). As discussed in Sec.4.3, no reliable predictions can be made
for the decays $\lamc\to\Xi^0K^+$ and $\xi0a\to\Sigma^+K^-$.

$$\vbox{\tabskip=0pt \offinterlineskip
  \def\tabrule{\noalign{\hrule}}
  \halign to \hsize{
  \strut
  \vrule #&\qquad#\hfil~~~ & \vrule # ~~~&\qquad #\hfil~~~ & \qquad #\hfil~~~ &
  \quad #\hfil~~~
  & \quad #\hfil~~~ & \quad #\hfil~~~   & \quad #\hfil~~~ & ~~~\qquad #\hfil~~
  &\quad # \hfil~~~ & ~~\qquad # \hfil~~
  &~~\vrule #\tabskip=0pt\cr
  \tabrule
  & & & & & & & & & & & & \cr
  & Reaction &  & $A^{\fac}$ & $A^{\pole}$ & $A^{\tot}$ & $B^{\fac}$
  & $B^{\pole}$
  & $ B^{\tot}$ & $\alpha$ &~~$\Gamma$~~ & $(Br)_{\rm theory}$ \quad &
   \cr
  & & & & & & & & & & & & \cr
  \tabrule
  & & & & & & & & & & & & \cr
  &$\lamc  \to p\bar{K}^0$ &
  & -5.73 & 3.91 & -1.82 & 14.33 & 3.23 & 17.56 & -0.49 & 0.63 & 1.20 & \cr
  &$\lamc \to \Lambda\pi^+$ & &
  -5.40 & 2.10 & -3.30 & 18.09 & -4.87 & 13.22 & -0.95 & 0.44 & 0.84 &  \cr
  &$\lamc\to \Sigma^0\pi^+$ & &
  0 & 2.20 & 2.20 & 0 & 14.63 & 14.63 & 0.78 & 0.36 & 0.68 &  \cr
  &$ \lamc\to \Sigma^+\pi^0$ & &
  0 & -2.20 & -2.20 & 0 & -14.63 & -14.63 & 0.78 & 0.36 & 0.68 &  \cr
  &$ \lamc\to \Xi^0 K^+$ & &
  0 & 0.10 & 0.10 & 0 & 4.26 & 4.26 & ----- & ----- & ----- &  \cr
  & & & & & & & & & & & & \cr
  &$ \xia\to \Sigma^+ \bar{K}^0$ & &
  -3.75 & 0.01 & -3.74 & 12.99 & -12.50 & 0.49 & -0.09 & 0.19 & 0.78 &  \cr
  &$ \xia\to \Xi^0 \pi^+$ & &
  6.99 & -0.06 & 6.93 & -24.90 & 14.24  &-10.66  &-0.77 & 0.89 & 3.65 &  \cr
  & & & & & & & & & & & & \cr
  &$ \xi0a\to \Lambda \bar{K}^0$ & &
  1.34 & 0.31  & 1.64  & -4.72 & -6.04 & -10.76 & -0.73 & 0.24 & 0.24 &  \cr
  &$ \xi0a\to \Sigma^0 \bar{K}^0$ & &
  2.65 & 0.12 & 2.77 & -9.19 & 6.39 & -2.80 & -0.59 & 0.12 & 0.12 &  \cr
  &$ \xi0a\to \Sigma^+ K^-$ & &
  0 & 0.16 & 0.16  & 0 & 3.53 & 3.53  & ----- & ----- & ----- &  \cr
  &$ \xi0a\to \Xi^0 \pi^0$ & &
  0 & 1.13 & 1.13 & 0 & -12.76 & -12.76 & -0.54 & 0.25 & 0.25 &  \cr
  &$ \xi0a\to \Xi^- \pi^+$ & &
  -7.18 & 1.55 & -5.63 & 25.10 & -3.80 & 21.30 & -0.99 & 1.12 & 1.12 & \cr
  & & & & & & & & & & & & \cr
  &$ \Omega_c^0 \to \Xi^0 \bar{K}^0$ & &
  1.72 & -0.06 & 1.66 & 2.75 & -10.20 & -7.45 & -0.93  & 0.13 &  & \cr
  & & & & & & & & & & & & \cr
  \tabrule
  }}$$

\vfill\eject

\hsize=9 in

Table II. Same as Table I except that predictions are made by current algebra.

$$\vbox{\tabskip=0pt \offinterlineskip
  \def\tabrule{\noalign{\hrule}}
  \halign to \hsize{
  \strut
  \vrule #&\qquad#\hfil~~~ & \vrule # ~~~&\qquad #\hfil~~~ & \qquad #\hfil~~~ &
  \quad #\hfil~~~
  & \quad #\hfil~~~ & \quad #\hfil~~~   & \quad #\hfil~~~ & ~\qquad #\hfil~~
  &\quad # \hfil~~~ & ~~\qquad # \hfil~~
  &~~\vrule #\tabskip=0pt\cr
  \tabrule
  & & & & & & & & & & & & \cr
  & Reaction &  & $A^{\fac}$ & $A^{\com}$ & $A^{\tot}$ & $B^{\fac}$
  & $B^{\pole}$
  & $ B^{\tot}$ & $\alpha$ &~~$\Gamma$~~ & $(Br)_{\rm theory}$ \quad &
   \cr
  & & & & & & & & & & & &  \cr
  \tabrule
  & & & & & & & & & & & &  \cr
  &$\lamc  \to p\bar{K}^0$ &
  & -5.73 & -4.44 & -10.17 & 14.33 & 2.10 & 16.43 & -0.90 & 1.82 & 3.46 & \cr
  &$\lamc \to \Lambda\pi^+$ & &
  -5.40 & 0 & -5.40 & 18.09 & -4.14 & 13.95 & -0.99 & 0.73 & 1.39 & \cr
  &$\lamc\to \Sigma^0\pi^+$ & &
  0 & -7.66 & -7.66 & 0 & 6.42 & 6.42 & -0.49 & 0.88 & 1.67 & \cr
  &$ \lamc\to \Sigma^+\pi^0$ & &
  0 & 7.66 & 7.66 & 0 & -6.42 & -6.42 &  -0.49 & 0.88 & 1.67 &  \cr
  &$ \lamc\to \Xi^0 K^+$ & &
  0 & -0.06 & -0.06 & 0 & -2.98 & -2.98 & ----- & ----- & ----- &  \cr
  & & & & & & & & & & & & \cr
  &$ \xia\to \Sigma^+ \bar{K}^0$ & &
  -3.75 & 4.47 & 0.72 & 12.99 & -12.50 & 0.49 & 0.43 & 0.01 & 0.03 &  \cr
  &$ \xia\to \Xi^0 \pi^+$ & &
  6.99 & -5.48 & 1.51 & -24.90 & 14.24  &-10.66  &-0.77 & 0.19 & 0.78 &  \cr
  & & & & & & & & & & & & \cr
  &$ \xi0a\to \Lambda \bar{K}^0$ & &
  1.34 & 5.55  & 6.89  & -4.72 & -6.78 & -11.50 & -0.88 & 0.89 & 0.89 &  \cr
  &$ \xi0a\to \Sigma^0 \bar{K}^0$ & &
  2.65 & -3.20 & -0.55 & -9.19 & 6.13 & -3.06 & 0.85 & 0.02 & 0.02 &  \cr
  &$ \xi0a\to \Sigma^+ K^-$ & &
  0 & 0.06 & 0.06  & 0 & -0.92 & -0.92  & ----- & ----- & ----- &  \cr
  &$ \xi0a\to \Xi^0 \pi^0$ & &
  0 & 7.76 & 7.76 & 0 & -12.20 & -12.20 & -0.78 & 1.12 & 1.12 &  \cr
  &$ \xi0a\to \Xi^- \pi^+$ & &
  -7.18 &5.49 & -1.69 & 25.10 & -2.74 & 22.36 & -0.47 & 0.74 & 0.74 & \cr
  & & & & & & & & & & & & \cr
  &$ \Omega_c^0 \to \Xi^0 \bar{K}^0$ & &
  1.72 & -11.05 & -9.33 & 2.75 & -9.54 & -6.79 & 0.44  & 0.98 &  & \cr
  & & & & & & & & & & & & \cr
  \tabrule
  }}$$

\vfill\eject

Table III. The predicted decay rates (in units of $10^{11}s^{-1}$) and the
$\alpha$ asymmetry parameter (in parentheses) for $B_c\rightarrow B+P$
decays in various models.

$$\vbox{\tabskip=0pt \offinterlineskip
  \def\tabrule{\noalign{\hrule}}
  \halign to \hsize{
  \strut
  \vrule #&\qquad#\hfil & ~\qquad \vrule # & ~\qquad #\hfil & \qquad #\hfil
  & ~~\qquad #\hfil & ~\qquad #\hfil & \qquad #\hfil
  & \qquad ~\vrule #\tabskip=0pt\cr
  \tabrule
  & & & & & & & & \cr
  & Reaction &&  Current algebra & Pole model & Xu and Kamal & K\"orner
  and Kr\"amer & Experiment  \quad & \cr
  & & &(this work) & (this work) & [15] & [16] & [22,24] & \cr
  & & & & & & & &  \cr
  \tabrule
  & & & & & & & &  \cr
  &$\lamc  \to p\bar{K}^0$ &
  & 1.82~(-0.90) & 0.63~(-0.49) & 0.60~(0.51) & input~(-1.0) & $0.84\pm 0.21$
  & \cr
  & $\lamc \to \Lambda\pi^+$ &
  & 0.73~(-0.99) & 0.44~(-0.95) & 0.81~(-0.67) & input~(-0.70) & $0.30\pm
  0.08~(-1.03\pm 0.29)$ &  \cr
  &$\lamc\to \Sigma^0\pi^+$ & &
  0.88~(-0.49) & 0.36~(0.78) & 0.17~( 0.92) & 0.16~(0.70) & $0.29\pm 0.14$ &
  \cr
  &$ \lamc\to \Sigma^+\pi^0$ & &
  0.88~(-0.49) & 0.36~(0.78) & 0.17~( 0.92) & 0.16~(0.70) &  & \cr
  &$ \lamc\to \Xi^0 K^+$ & &
  ----- & ----- & 0.05~(0) & 0.13~(0) & $0.17\pm 0.04$ & \cr
  & & & & & & & &  \cr
  &$ \xia\to \Sigma^+ \bar{K}^0$ & &
  0.01~(0.43) & 0.19~(-0.09) & 0.10~(0.24)& 1.46~(-1.0) &  &\cr
  &$ \xia\to \Xi^0 \pi^+$ & &
  0.19~(-0.77) & 0.89~(-0.77) & 0.76~(-0.81) & 0.80~(-0.78) & &\cr
  & & & & & & & & \cr
  &$ \xi0a\to \Lambda \bar{K}^0$ & &
  0.89~(-0.88) & 0.24~(-0.73) & 0.33~(1.0) & 0.11~(-0.76) & &\cr
  &$ \xi0a\to \Sigma^0 \bar{K}^0$ & &
  0.02~(0.85) & 0.12~(-0.59)& 0.09~(-0.99) & 1.05~(-0.96) & &\cr
  &$ \xi0a\to \Sigma^+ K^-$ & &
  ----- & ----- & 0.11~(0) & 0.11~(0) && \cr
  &$ \xi0a\to \Xi^0 \pi^0$ & &
  1.12~(-0.78) & 0.25~(-0.54) & 0.50~(0.92) & 0.03~(0.92) &   &\cr
  &$ \xi0a\to \Xi^- \pi^+$ & &
  0.74~(-0.47) & 1.12~(-0.99) & 1.55~(-0.38) & 0.93~(-0.38) && \cr
  & & & & & & & & \cr
  &$ \Omega_c^0 \to \Xi^0 \bar{K}^0$ & &
  0.98~(0.44) & 0.13~(-0.93) & & 1.75~(0.51) & & \cr
  & & & & & & & &  \cr
  \tabrule
  }}$$

\vfill\eject

\end